# Electrostatic shielding effect of ground-state energy of metallic and nonmetallic elements

Maolin Bo, [1]* Hanze Li,[2] Zhihong Wang, [1] Yunqian Zhong, [1] Yao Chuang, [1]

ZhongKai Huang[1]

[1]Key Laboratory of Extraordinary Bond Engineering and Advanced Materials Technology (EBEAM) of Chongqing, Yangtze Normal University, Chongqing 408100, China

[2]School of Physics and Optoelectronics, Xiangtan University, Hunan 411105, China

*Corresponding Author: E-mail: bmlwd@yznu.edu.cn (Maolin Bo)

**Abstract**

The ground-state energy is crucial for studying the properties of a material. In this study, we computed both the Hartree–Fock and random phase approximations of the ground state energy of metals and nonmetals. Considering the effect of the electrostatic shielding potential, we utilized the Thomas–Fermi dielectric function to obtain the Thomas–Fermi formula for the total potential energy. We calculated the total potential energy of the elements using the Wigner and Hedin–Lundqvist correlation energies, considering the changes in them after the electrostatic shielding effect. The exchange correlation potential, including electrostatic shielding effect, can be used in the measurement of scanning impedance microscopy experiments. Furthermore, we consider the dynamic process with respect to the change in the potential energy.

# 1. Introduction

The ground state refers to the state in which an atom is at its lowest energy level and the electrons move in the orbit closest to the nucleus[1, 2]. At this time, the electrons are in a stable equilibrium state, given that the lower the energy, the more stable the system. The distribution of electrons in atomic orbitals should give rise to the atom being in the lowest energy state possible, which is known as the principle of minimum energy. The properties of materials are determined by the ground state and excited state of the electrons, and various forms of matter are largely the manifestation of the electronic ground state. Additionally, the stable structure of solids is usually classified according to the ground state, because the lowest energy electronic state determines the bonding of atoms, or the spatial structure of the nucleus. The ground state of electrons is therefore very important for the electronic structure. To better study the properties of materials, it is necessary to make accurate theoretical predictions and calculations of the ground-state energy. Therefore, the influence of the electrostatic shielding effect is crucial to correlation energy calculations[3]. In addition, the influence of the external field on the change in potential energy is involved in the chemical reaction process. Consideration of the temporal change in the potential energy under external field stimulation is a key step.

In this study, we used the Hartree–Fock approximation[4] (HFA) and the random phase approximation (RPA) to calculate the ground-state energy, obtain the relationship between this energy and the radius, and compare and analyze the calculation results. Further, we investigated changes due to the electrostatic shielding effect[5, 6]. After accounting for the electrostatic shielding effects, the exchange correlation potential successfully included the effects of electrostatic shielding[7-9]. Scanning impedance microscopy (SIM) is a powerful technique for characterizing the various properties of carrier motion in materials and devices under an oscillating electric field[10, 11]. The exchange correlation potential including electrostatic shielding effect can be used in the measurement of SIM experiments[12, 13]. Furthermore, we considered the dynamic process with respect to the change in potential energy.

## 2. Principles

### 2.1 Ground-state energy

#### 2.1.1 Ground-state energy for Hartree–Fock approximation (HFA)

The ground-state energy obtained by the HFA is as follows:

$$\frac{1}{N} E_0(HFA) = \frac{2.21}{r_s^2} - \frac{0.916}{r_s} \quad [R_y],$$

(1)

where energy is in Rydberg units $[R_y]$, such that:

$$1[Ry] \equiv \frac{me^4}{2\hbar^2} = \frac{e^2}{2a_H} = 13.6 eV$$

(2)

where $r_s$ is the atomic radius, $h$ is Planck's constant and is expressed as $\hbar = \frac{h}{2\pi}$, $a_H$ is the Bohr radius, and $e$ is the electron charge. The HFA only considers the exchange between electrons and dismisses the correlation energy correction of many-body systems.

#### 2.1.2 Ground-state energy of random phase approximation (RPA)

M. Gell-Mann and K.A. Bruckner[14] obtained the correlation energy correction, $\varepsilon_c$, of high-density electronic systems through complex calculations as follows:

$$\varepsilon_c = 0.0622 \ln r_s - 0.094 \quad [Ry]$$

(3)

where $r_s$ is the atomic radius and the energy is in Rydberg units $[Ry]$. Adding this formula to the ground-state energy equation in **Eq. 1** yields the average ground-state energy for each lattice point:

$$\frac{1}{N} E_0 = \frac{2.21}{r_s^2} - \frac{0.916}{r_s} - 0.094 + 0.0622 \ln r_s + o(r_s \ln r_s) \quad [Ry],$$

(4)

where $o(r_s \ln r_s)$ is a higher-order infinitesimal function and is not included in the calculation.

**2.1.2 Electrostatic shielding potential**

A static external disturbance corresponds to a frequency of $\omega=0$, and the Thomas–Fermi dielectric function[15, 16] is as follows:

$$\varepsilon_1(q,0) \approx 1+\left(\frac{6\pi N e^2}{E_F}\right)\frac{1}{q^2} \equiv 1+\frac{\lambda^2}{q^2},$$

(5)

where $N$ is the number of lattice points, $e$ is the electron charge, $E_F$ is the Fermi energy, $q$ is the wave vector, and $\lambda$ represents the shielding parameters. Using this formula to investigate the case in which there is a negative charge at the origin, the following Thomas–Fermi formula for the total potential energy can be obtained:

$$V_{TF}(r) = \sum_q \frac{4\pi e^2}{q^2+\lambda^2} e^{iq\cdot r} = \frac{e^2}{r_s} e^{-\lambda r_s},$$

(6)

where $V_{TF}$ is the electrostatic shielding potential. This equation demonstrates that a $(-e)$ charge located at the origin will be shielded within a distance of $\lambda^{-1}$, and the shielding effect of the electronic system will give rise to a Coulomb interaction between the electrons in the metal of $(e^2/r)$ being converted into a Yukawa type interaction. The effect is then represented by $(e^2/r)\exp(-\lambda r)$, where $\lambda$ and $q_c$ are of the same magnitude, and $q_c$ is the upper limit of the plasmon wavenumber. Here, we define the scaling factor $L$ of electrostatic shielding. Considering the effect of electrostatic shielding for the potential function, we can obtain the following:

$$V_i = V_0 e^{-\lambda r_s},$$

(7)

where $V_0$ is the initial potential function, and $V_i$ is the potential function after considering electrostatic shielding.

## 2.2 Exchange Correlation Approximation

### 2.2.1 Wigner correlation energy approximation

Proposed by E. Wigner, the D. Pines[17] corrected correlation energy is:

$$\varepsilon_c = -\frac{0.88}{r_s + 7.79},$$

(8)

and the resulting correlation potential is:

$$V_c(r_s) = -0.88 \frac{2/3 r_s + 7.79}{(r_s + 7.79)^2},$$

(9)

where $r_s$ represents the atomic radius, including the two approximations described below.

### 2.2.2 Hedin–Lundqvist correlation energy approximation

This form of correlation energy was proposed by L. Hedin and B. Lundqvist[3], where the correlation energy formula is as follows:

$$\varepsilon_c(r_s) = -\frac{0.045}{2}\left\{1 + (r_s/21)^2 \ln(1 + 21/r_s) + \frac{r_s}{42} - (r_s/21)^2 - \frac{1}{3}\right\}.$$

(10)

The correlation potential is:

$$V_c(r_s) = -\frac{0.045}{2} \ln\left(1 + \frac{21}{r_s}\right).$$

(11)

### 2.2.3 Ceperley–Alder exchange-correlation potential approximation

The exchange correlation energy was proposed by T.P Perdew and A.Zunger[18], where:

$$\varepsilon_x(r_s) = -0.9164/r_s,$$

$$\varepsilon_c(r_s) = \begin{cases} -0.2846/\left(1+1.0529\sqrt{r_s}+0.3334r_s\right) & (r_s \geq 1) \\ -0.0960+0.0622\ln r_s -0.0232r_s +0.0040r_s \ln r_s & (r_s \leq 1) \end{cases}$$

(12)

(13)

This is currently the most used approximately method[19, 20]. The exchange association energy can be composed of two parts, i.e., exchange and association, such that

$$\varepsilon_{xc} = \varepsilon_x + \varepsilon_c,$$

(14)

where $\varepsilon_x$ represents the exchange energy and $\varepsilon_c$ represents the correlation energy. Two commonly used exchange-correlation energy approximations are described below.

**2.3 Phase operator and coherent state**

The particle fluctuation of coherent state is

$$\begin{cases} \langle \alpha|\hat{N}|\alpha\rangle = |\alpha|^2 \\ \langle \alpha|\hat{N}^2|\alpha\rangle = \langle \alpha|\hat{a}^\dagger \hat{a} \hat{a}^\dagger \hat{a}|\alpha\rangle = |\alpha|^4 + |\alpha|^2 \end{cases},$$

(15)

where $\hat{N}$ is particle number operator that satisfies $\hat{N}=\hat{a}^\dagger \hat{a}$. The fluctuation of coherent state ($\alpha = |\alpha|e^{i\phi}$) and relative uncertainty are $\langle(\Delta\hat{N})^2\rangle = |\alpha|^2 = \langle\hat{N}\rangle$ and $\langle\Delta\hat{N}\rangle/\langle\hat{N}\rangle = \dfrac{1}{|\alpha|}$, respectively. The expected value of the phase operator satisfied

$$\langle\alpha|\hat{C}|\alpha\rangle = \frac{1}{2}\exp(-|\alpha|^2)\sum_{n=0}^{\infty}\frac{(\alpha^*)^{n+1}\alpha^n + (\alpha^*)^n \alpha^{n+1}}{\sqrt{(n+1)!n!}}$$

$$= |\alpha|\cos\phi\exp(-|\alpha|^2)\sum_{n=0}^{\infty}\frac{|\alpha|^{2n}}{n!\sqrt{n+1}}.$$

$$(\hat{C} \equiv \frac{1}{2}\left[\left(\hat{N}+1\right)^{-1/2}\hat{a} + \hat{a}^\dagger\left(\hat{N}+1\right)^{-1/2}\right])$$

(16)

where phase operator is proportional to $\cos\phi$ and $\phi$ is the argument of coherent state $\alpha$ eigenvalue. Similarly, the expected value of $\hat{C}^2$ is

$$\left\langle \alpha \left| \hat{C}^2 \right| \alpha \right\rangle = \frac{1}{2} - \frac{1}{4}\exp\left(-|\alpha|^2\right)$$
$$+ |\alpha|^2 \left(\cos^2\phi - \frac{1}{2}\right)\exp\left(-|\alpha|^2\right)\sum_{n=0}^{\infty}\frac{|\alpha|^{2n}}{n!\sqrt{(n+1)(n+2)}}.$$

(17)

Generally, the simplification of the analytical solution form should not be obtained from **Eq. 17**; however, the progressive form is the better choice, which is

$$\sum_{n=0}^{\infty}\frac{|\alpha|^{2n}}{n!\sqrt{n+1}} = \frac{\exp(|\alpha|^2)}{|\alpha|}\left(1 - \frac{1}{8|\alpha|^2} + ...\right),$$

$$\sum_{n=0}^{\infty}\frac{|\alpha|^{2n}}{n!\sqrt{(n+1)(n+2)}} = \frac{\exp(|\alpha|^2)}{|\alpha|}\left(1 - \frac{1}{2|\alpha|^2} + ...\right),$$

(18)

Thus,

$$\left\langle \alpha \left| \hat{C} \right| \alpha \right\rangle = \cos\phi\left(1 - \frac{1}{8|\alpha|^2} + ...\right) \simeq \cos\phi, \quad |\alpha|^2 \gg 1$$

$$\left\langle \alpha \left| \hat{C}^2 \right| \alpha \right\rangle = \cos\phi - \frac{\cos^2\phi - 1/2}{2|\alpha|^2} + ...,$$

(19)

Finally, we obtain the uncertainty relation of phase operator as

$$\left\langle \Delta \hat{C}^2 \right\rangle = \left\langle \hat{C}^2 \right\rangle - \left\langle \hat{C} \right\rangle^2 \simeq \frac{\sin^2\phi}{8|\alpha|^2}, \quad |\alpha|^2 \gg 1.$$

(20)

Similarly, the fluctuation of the other phase operator $\hat{S}$ is

$$\left\langle \alpha \left| \hat{S} \right| \alpha \right\rangle = \sin\phi \left( 1 - \frac{1}{8|\alpha|^2} + ... \right)$$

$$(\hat{S} \equiv \frac{1}{2i}\left[ \left(\hat{N}+1\right)^{-1/2} \hat{a} - \hat{a}^\dagger \left(\hat{N}+1\right)^{-1/2} \right]).$$

(21)

When $|\alpha|^2 \gg 1$, we have

$$\left\langle \alpha \left| \hat{S} \right| \alpha \right\rangle \simeq \sin\phi.$$

(22)

As observed in the form of $\left\langle \left(\Delta\hat{N}\right)^2 \right\rangle / \left\langle \hat{N} \right\rangle$, with the increase in $|\alpha|^2$, the expected values of the amplitude and phase of the coherent state tend to become more accurate.

## 2.4 Dynamic process of potential energy

Considering the dynamic process of potential function:

$$\begin{cases} V_i(\vec{r},t) = (1+f(t))V_0(\vec{r}) \\ \mathcal{L}(f(t)) = F(s) = \int_{-\infty}^{+\infty} f(t)e^{-st}dt \\ f(t) = \mathcal{L}^{-1}(F(s)), \hat{S}_t = e^{-iEt/\hbar} = e^{-iwt} \end{cases}$$

(23)

$\mathcal{L}$ is Laplace transform. $\hat{S}_t$ is an operator. The $s=\kappa+jw$ is a complex number. When $\kappa$ is 0, the function $f(t)$ is Fourier transform.

Consider the effects of external fields: $f(t)=A(t)$. For first-order systems with time response:

1) differential equation:

$$\frac{dA(t)}{dt} + aA(t) = au(t).$$

(26)

2) transfer function:

$$G(s) = \frac{A'(s)}{U(s)} = \frac{a}{s+a}.$$

3) unit step function:

$$u(t) = \begin{cases} 1, & t \geq 0 \\ 0, & t < 0 \end{cases}.$$

(27)

4) time-domain expression of unit step response:

$$A(t) = 1 - e^{-at}$$

(28)

For time responsive second-order systems:

1) Differential equation:

$$\frac{d^2 A(t)}{dt^2} + 2\zeta\omega_n \frac{dA(t)}{dt} + \omega_n^2 A(t) = \omega_n^2 u(t)$$

(29)

2) Transfer function:

$$G(s) = \frac{A'(s)}{U(s)} = \frac{\omega_n^2}{s^2 + 2\zeta\omega_n s + \omega_n^2}$$

(30)

3) Unit step function:

$$u(t) = \begin{cases} 1, & t \geq 0 \\ 0, & t < 0 \end{cases}$$

(31)

4) Time-domain expression of unit step response:

(1) Underdamped system $(0 < \zeta < 1)$:

$$A(t) = 1 - e^{-\zeta\omega_n t}\left[\cos\omega_d t + \frac{\zeta}{\sqrt{1-\zeta^2}}\sin\omega_d t\right].$$

(32)

(2) Undamped system $(\zeta = 0)$:

$$A(t) = 1 - \cos\omega_n t.$$

(33)

(34)

(3) Critical damping system $(\zeta = 1)$:

$$A(t) = 1 - e^{-\omega_n t}(1 + \omega_n t).$$

(35)

(4) Damping system $(\zeta > 1)$:

$$A(t) = 1 - \frac{1}{2\sqrt{\zeta^2-1}\left(\zeta - \sqrt{\zeta^2-1}\right)} e^{\left(-\zeta\omega_n + \omega_n\sqrt{\zeta^2-1}\right)t} + \frac{1}{2\sqrt{\zeta^2-1}\left(\zeta + \sqrt{\zeta^2-1}\right)} e^{\left(-\zeta\omega_n - \omega_n\sqrt{\zeta^2-1}\right)t}.$$

(36)

If $A(t)$ is a second-order time response function, the Bessel equation ($n$ is a real constant) can be used to calculate the time response function:

$$t^2 \frac{d^2 A(t)}{dt^2} + t \frac{dA(t)}{dt} + (t^2 - n^2)A(t) = 0.$$

(37)

Bessel function of the first kind (represented by $J_n(t)$ and $J_{-n}(t)$) constitutes a group of fundamental solutions $J_n(t)$ equation, which is defined as follows:

$$J_n(t) = \left(\frac{t}{2}\right)^n \sum_{(k=0)}^{\infty} \frac{\left(\frac{-t^2}{2}\right)^k}{k!\Gamma(n+k+1)} \quad \left(\Gamma(n+k+1) = \int_0^{\infty} t^{n+k} e^{-n} dt\right).$$

(38)

The Bessel function of the second kind (expressed as $Y_n(t)$) constitutes another solution of the Bessel equation, which is linearly independent of $J_n(t)$. $Y_n(t)$ is defined as follows:

$$Y_n(t) = \frac{J_n(t)\cos n\pi - J_{-n}(t)}{\sin(n\pi)}.$$

(39)

The Bessel functions are related to the Hankel functions, also known as the third kind of Bessel functions:

$$\begin{cases} H_n^1(t) = J_n(t) + iY_n(t) \\ H_n^2(t) = J_n(t) - iY_n(t) \end{cases},$$

(40)

where $H_n^k(t)$ is besselh, $J_n(t)$ is besselj, and $Y_n(t)$ is bessely. The Hankel function also constitutes a group of basic solutions of the Bessel equation.

## 3. Results and discussion

### 3.2.1 Ground-state energy

We use **Eqs. 1** and **4** to calculate the ground-state energies by the HFA and RPA, and obtain data listed in **Tables 1 and 2** from the ground-state energy formula. Next, we plotted the ground-state energy and atomic radius. The relationship between the two methods and a comparison of both when calculating the ground-state energy are illustrated in **Fig. 1 (a)(b)**. **Tables 1** and **2** present the ground-state energy results calculated by the HFA and RPA, respectively. In **Tables 1** and **2**, the unit of radius $r_s$ is Angstrom $\dot{A}$, and the unit of ground-state energy, exchange correlation energy, and potential energy is electronvolt (eV).

For ease of discussion, we divided the analysis into metallic and nonmetallic elements. The red and green lines represent the ground-state energies calculated by the HFA and RPA, respectively. As indicated in the tables, to compare these two calculation methods, we calculate the difference between the calculated results of the ground-state energy of both methods. However, the difference between the two ground-state energy values reduced as the atomic radius increased. By comparing the difference between the two equations, we observed that the ground-state energy of each lattice point was represented by the RPA, which is based on the HFA and considers the interactions between electrons, i.e., $\varepsilon_c = 0.0622 \ln r_s - 0.094$.

As the contribution of the logarithmic term is negative, the correlation energy $\varepsilon_c$ is always negative. This highlights how including the RPA calculation in the electronic system. The disadvantage of HFA is that the ground-state energy obtained through it is too small, which is eliminated by the inclusion of RPA. As $r_s$ gradually increases, the value of the logarithmic function also increases. We inferred that the correlation energy also slowly increases. Hence, the difference between the ground-state energies of the HFA and RPA declines. In addition, Tables 1 and 2 also present the results of the exchange energy approximated by the Ceperley–Alder exchange correlation. The Ceperley–Alder correlation energy is larger than that of the RPA. However, the exchange correlation energy of the local density approximation which were calculated using the Monte-Carlo method[15].

**3.2.2 Electrostatic shielding potential and exchange correlation potential**

This paper adopts the Thomas–Fermi formula of the total potential energy[21]:

$$V_{TF}(r_s) = \frac{e^2}{r_s} e^{-\lambda r_s},$$

(41)

where the shielding parameter $\lambda$ was varied as 0.1, 0.3, 0.5, 0.7, and 1 for the calculation. **Fig. 3** illustrates a comparison of the relationship between the electrostatic shielding potential and the atomic radius when different $\lambda$ were used. As observed in **Table 3** and **Fig. 3**, as $r_s$ increases, the electrostatic shielding potential decreases, and the closer the shielding parameter value to 1. This implies a smaller electrostatic shielding potential.

**Eqs. 9 and 11** are the Wigner and Hedin–Lundqvist correlation energy approximations, respectively. When using the $r_s$ data to calculate the correlation potential, the two approximations do not consider the electrostatic shielding potential. If the Wigner correlation energy is approximated by considering the scaling factor *L* (**Eqs**. **7** and **8)** of the electrostatic shielding effect, we obtain the following:

$$V_c(r_s) = -0.88 \frac{2/3 r_s + 7.79}{(r_s + 7.79)^2} e^{-\lambda r_s},$$

(42)

and the corresponding approximation of the Hedin–Lundqvist correlation is

$$V_c(r_s) = -\frac{0.045}{2} \ln\left(1 + \frac{21}{r_s}\right) e^{-\lambda r_s}.$$

(43)

The masking parameter $\lambda$ was varied by 0.1, 0.3, 0.5, 0.7, and 1, respectively. The results are presented in **Tables 4** and **5**. The two potential functions are compared in **Figs. 4** and **5** (where no electrostatic shielding effect is considered). As observed in **Tables 4** and **5**, the correlation potential gradually increases as the atomic radius increases.

**Figs. 4** and **5** display the results of the Wigner and Hedin–Lundqvist correlation energy approximations after considering the electrostatic shielding effect. Evidently, the correlation potential from the Wigner approximation was smaller than that from the Hedin–Lundqvist approximation. The orange dotted lines in the two figures represent the correlation potential without considering the electrostatic shielding effect. After considering the electrostatic shielding effect, the results of the two correlation potentials increased, and as the value of the shielding parameter increased, the correlation potential also increased.

Considering the energy scaling factor $L$ for the Wigner[22, 23] and Hedin–Lundqvist correlation energies[24, 25], we obtain the following:

$$L = \left|\frac{V_i(\vec{r_s})}{V_0(\vec{r_s})}\right| = e^{-\lambda_i r_s}.$$

(44)

Next, we derive the following:

$$V_i(\vec{r_s}) = V_0(\vec{r_s}) e^{-\lambda_A r_s}$$

(45)

$$V_i'(\vec{r_s}) = -V_0(\vec{r_s}) e^{-\lambda_B r_s}$$

(46)

The value of $\lambda_A$ in **Eq. 43** is related to the electrostatic shielding effect. The value of

$\lambda_A$ is inversely proportional to this effect; therefore, the electronic polarization decreases as $\lambda_A$ increases, and the potential energy decreases. The $\lambda_B$ in **Eq. 44** is related to the antibonding potential energy. An increase in $\lambda_B$ will simultaneously decrease the antibonding potential energy. The energy scales $\lambda_B$ and $\lambda_B$ apply only to the comparison of the same composition material. The formula is linked to the BBC model, which can be found in **supporting materials**.

To represent the change in the potential-energy surface and phase operator, we use the following formula:

$$V_i = \frac{1}{2}(V_i(\vec{r_s})+V_i'(\vec{r_s}))-\frac{1}{4}(V_i(\vec{r_s})-V_i'(\vec{r_s}))(\sin x + \cos y).$$

(47)

If we dismiss the antibonding potential energy, we derive the following formula:

$$V_i = \frac{1}{2}V_i(\vec{r_s})-\frac{1}{4}V_i(\vec{r_s})(\sin x + \cos y).$$

(48)

Here, $x$ and $y$ are the phase positions of the wave function. A and B represent the amplitudes of the wave. **Figs. 6 and 7** depict the Wigner and Hedin–Lundqvist correlation energies of the potential surfaces of Li, P, and B. The potential energy surface can be measured by SIM. The three-dimensional potential energy surface and phase operator can be calculated as follows:

$$V_i(\vec{r_s})=V_0(\vec{r_s})(e^{-\lambda_x r_s}\cos x + e^{-\lambda_y r_s}\cos y + e^{-\lambda_z r_s}\cos z)$$

(49)

$x$, $y$ and $z$ are the phase positions of the wave function.

According to **Eqs. 25**, **37,** and **38**, the dynamic formula of the besselj time-dependent potential function is:

$$V_i(\vec{r_s}, t) = (1 + J_n(t))V_i(\vec{r_s}).$$

(50)

**Fig. 8** displays the besselj (n=1) time-dependent potential function of Li. The dimensionless time is $\tau = t/t_0$, and the characteristic time is $t = \sqrt{mR_{ij}^2/\varepsilon}$. $R_{ij}$ is the atomic distance and $m$ is the atomic mass. $\varepsilon$ is the depth of the potential well, usually is $1*10^{-19}$ J. The general characteristic time $t$ is $1*10^{-13}$ s, that is, 100 fs.

## 4. Conclusions

The HFA and RPA are well-established and effective methods for calculating ground-state energy. In this study, we used them to calculate the ground-state energies of metals and nonmetals and compared the difference between them. The ground-state energy of the HFA was always higher than that of the RPA. However, this difference narrowed with the increase in the atomic radius because the RPA considered the interaction between electrons. Further, with the HFA, the correlation energy gradually increased. However, it remained negative, so the difference between the ground-state energies of the HFA and RPA decreased. Additionally, using the Thomas–Fermi dielectric function of the total potential energy to calculate the electrostatic shielding potential with different shielding parameters revealed that the potential decreased as the atomic radius increased. As the shielding parameter became closer to 1, the potential dwindled. We calculated the correlation potential approximated by the Wigner and Hedin–Lundqvist correlation energies. The correlation potential calculated by the Wigner correlation energy approximation is smaller than that calculated by the Hedin–Lundqvist correlation energy approximation. The approximations considered the scaling factor $L$ after the electrostatic shielding effect, and the two correlation potentials increased. Further, with an increase in the shielding parameter, the correlation potential also increased. By accounting for the scaling factor $L$ of the electrostatic shielding effect, the exchange correlation potential could successfully include the effects of shielding. Furthermore, the dynamic process of potential energy change can be obtained through the time response function. The potential energy surface can be computed by SIM.

**Figure and Table Captions**

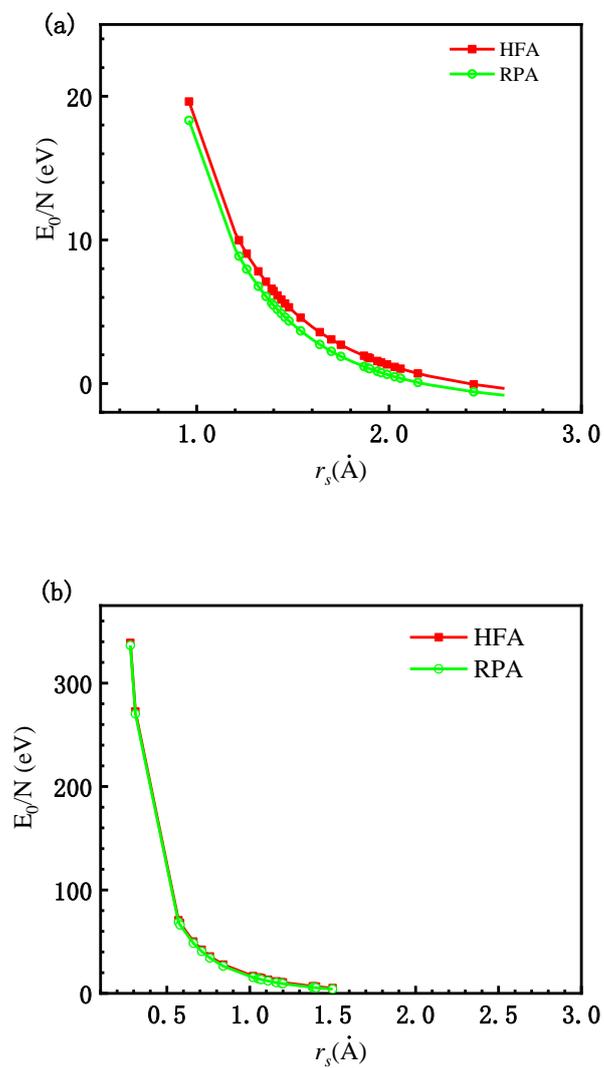

**Fig. 1** Comparison of the ground-state energies of the Hartree–Fock (HFA) and random phase approximation (RPA). (a) Metals and (b) nonmetals.

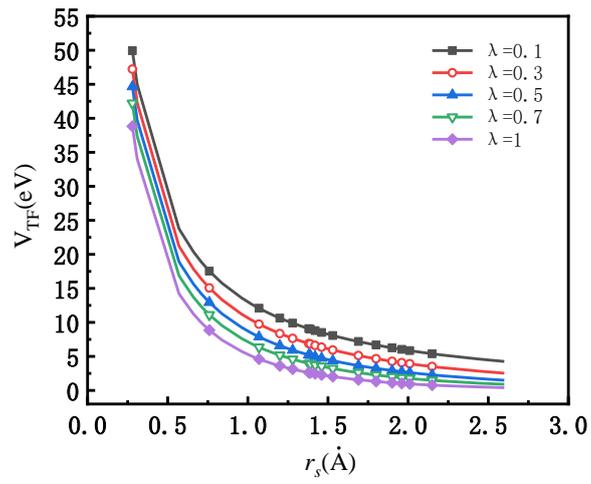

**Fig. 2** Electrostatic shielding potential vs. atomic radius $r_s$ for shielding parameter $\lambda$.

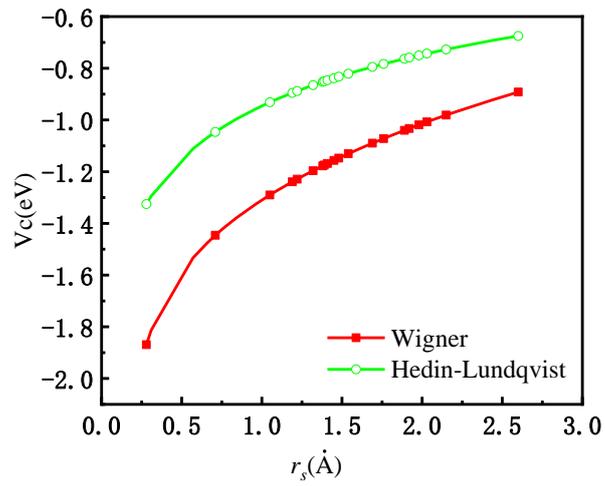

**Fig. 3** Comparison of Wigner and Hedin–Lundqvist correlation potentials with atomic radius $r_s$.

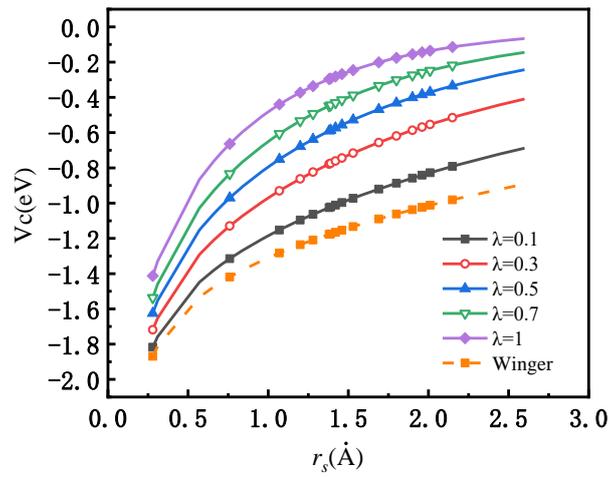

**Fig. 4** Wigner correlation potential vs. atomic radius considering electrostatic shielding effect.

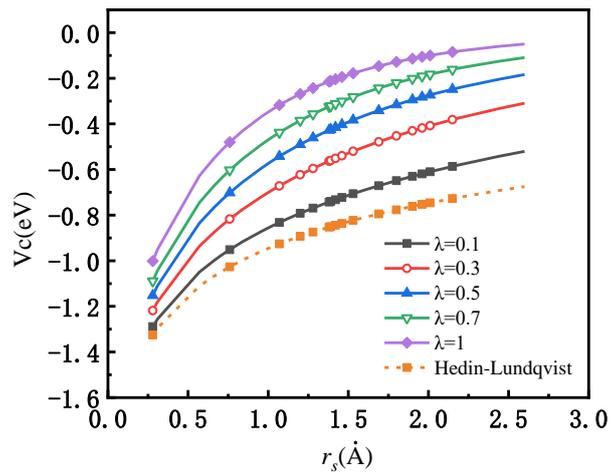

**Fig. 5** Hedin–Lundqvist correlation potential vs. atomic radius considering electrostatic shielding effect.

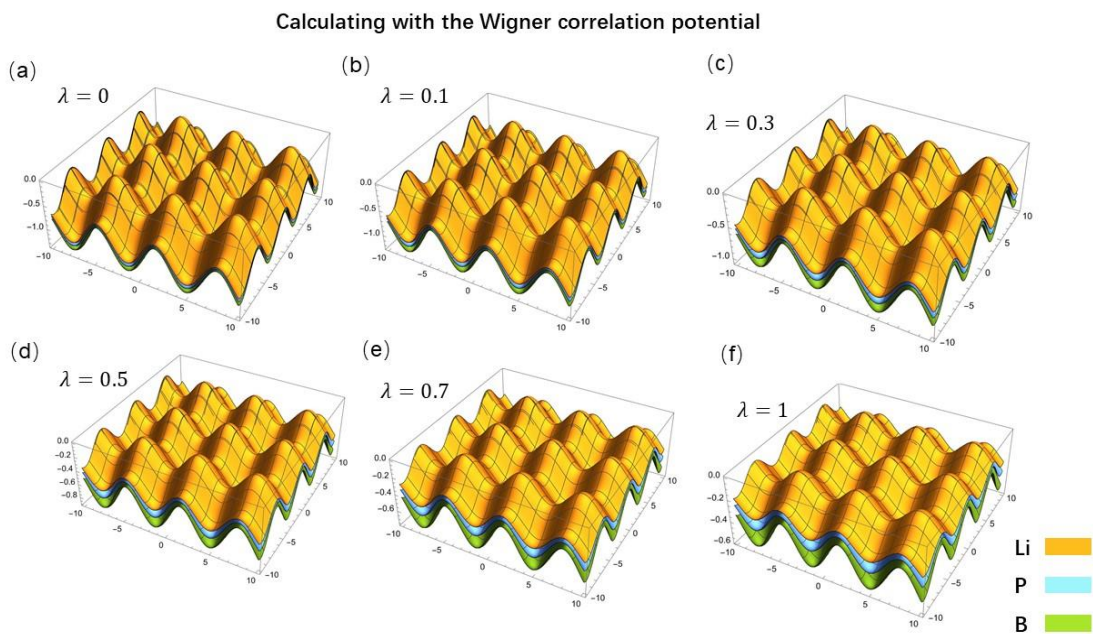

**Fig. 6** Comparison of Wigner correlation potential surfaces considering electrostatic shielding effect ($\lambda = 0, 0.1, 0.3, 0.5, 0.7, 1$) of Li, P, and B.

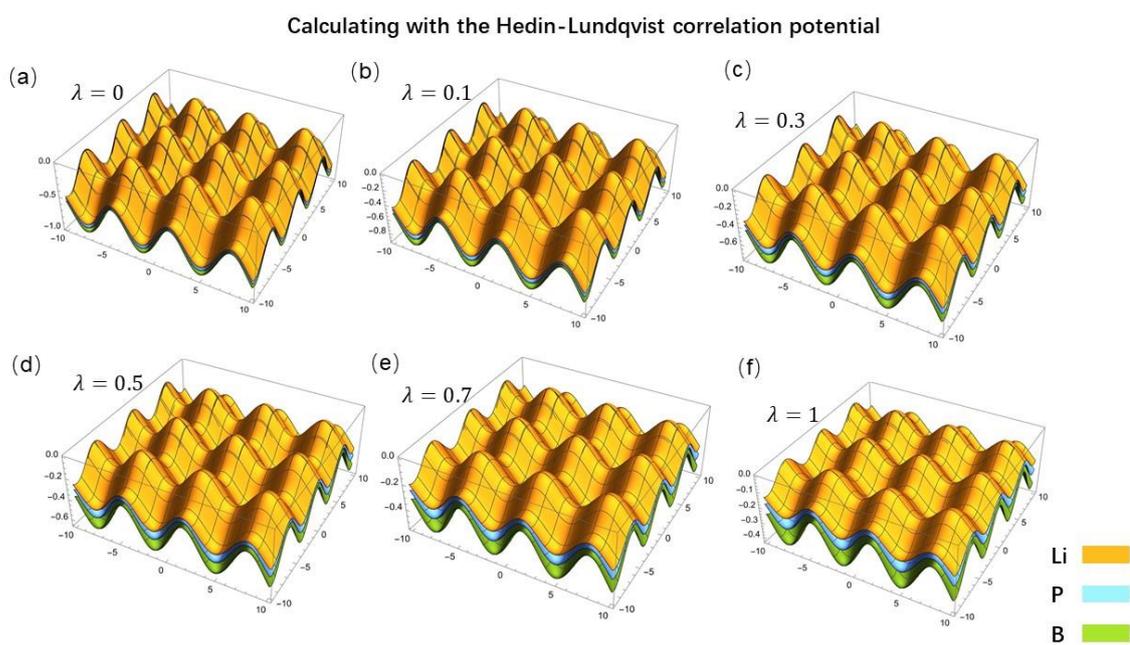

**Fig. 7** Comparison of Hedin–Lundqvist correlation potential surfaces considering electrostatic shielding effect ($\lambda = 0, 0.1, 0.3, 0.5, 0.7, 1$) of Li, P, and B.

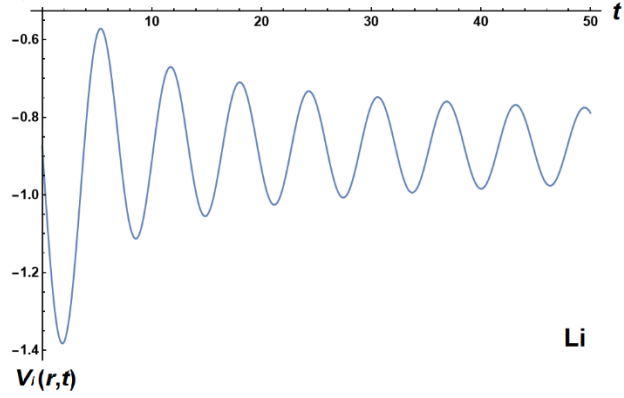

**Fig. 8** Besselj time-dependent potential function of Li.

**Table 1** Ground-state energies of metal elements using HFA and RPA. $\varepsilon_x$ and $\varepsilon_c$ represent the exchange and correlation energies, respectively.

| element | $r_s$(Å) | HFA | RPA | PRA -H FA | $\varepsilon_x$ | $\varepsilon_c$ |
|---|---|---|---|---|---|---|
| **Be** | 0.96 | 19.6362 | 18.3232 | -1.3129 | -12.9823 | -1.6452 |
| **Ge** | 1.20 | 10.4909 | 9.3667 | -1.1242 | -10.3859 | -1.8734 |
| **Al** | 1.21 | 10.2331 | 9.1160 | -1.1172 | -10.3000 | -1.8679 |
| **Zn** | 1.22 | 9.9823 | 8.8722 | -1.1102 | -10.2156 | -1.8624 |
| **Ga** | 1.22 | 9.9823 | 8.8722 | -1.1102 | -10.2156 | -1.8624 |
| **Ni** | 1.24 | 9.5009 | 8.4045 | -1.0964 | -10.0508 | -1.8516 |
| **Co** | 1.26 | 9.0447 | 7.9618 | -1.0829 | -9.8913 | -1.8410 |
| **Li** | 1.28 | 8.6122 | 7.5427 | -1.0696 | -9.7368 | -1.8305 |
| **Fe** | 1.32 | 7.8122 | 6.7686 | -1.0435 | -9.4417 | -1.8099 |
| **Cu** | 1.32 | 7.8122 | 6.7686 | -1.0435 | -9.4417 | -1.8099 |
| **Hg** | 1.32 | 7.8122 | 6.7686 | -1.0435 | -9.4417 | -1.8099 |
| **Pt** | 1.36 | 7.0900 | 6.0717 | -1.0183 | -9.1640 | -1.7900 |
| **Au** | 1.36 | 7.0900 | 6.0717 | -1.0183 | -9.1640 | -1.7900 |
| **Cr** | 1.39 | 6.5938 | 5.5940 | -0.9998 | -8.9662 | -1.7753 |
| **Mn** | 1.39 | 6.5938 | 5.5940 | -0.9998 | -8.9662 | -1.7753 |
| **Pd** | 1.39 | 6.5938 | 5.5940 | -0.9998 | -8.9662 | -1.7753 |
| **Sn** | 1.39 | 6.5938 | 5.5940 | -0.9998 | -8.9662 | -1.7753 |
| **Sb** | 1.39 | 6.5938 | 5.5940 | -0.9998 | -8.9662 | -1.7753 |
| **Po** | 1.40 | 6.4364 | 5.4426 | -0.9938 | -8.9022 | -1.7705 |
| **Mg** | 1.41 | 6.2828 | 5.2950 | -0.9878 | -8.8390 | -1.7658 |
| **Ir** | 1.41 | 6.2828 | 5.2950 | -0.9878 | -8.8390 | -1.7658 |
| **Rh** | 1.42 | 6.1328 | 5.1510 | -0.9818 | -8.7768 | -1.7610 |
| **In** | 1.42 | 6.1328 | 5.1510 | -0.9818 | -8.7768 | -1.7610 |
| **Cd** | 1.44 | 5.8435 | 4.8735 | -0.9699 | -8.6549 | -1.7517 |
| **Os** | 1.44 | 5.8435 | 4.8735 | -0.9699 | -8.6549 | -1.7517 |
| **Ag** | 1.45 | 5.7039 | 4.7398 | -0.9641 | -8.5952 | -1.7470 |
| **Ti** | 1.45 | 5.7039 | 4.7398 | -0.9641 | -8.5952 | -1.7470 |
| **Ru** | 1.46 | 5.5676 | 4.6093 | -0.9583 | -8.5363 | -1.7424 |
| **Pb** | 1.46 | 5.5676 | 4.6093 | -0.9583 | -8.5363 | -1.7424 |
| **Tc** | 1.47 | 5.4345 | 4.4820 | -0.9525 | -8.4783 | -1.7379 |
| **Bi** | 1.48 | 5.3044 | 4.3576 | -0.9468 | -8.4210 | -1.7333 |
| **Re** | 1.51 | 4.9318 | 4.0020 | -0.9298 | -8.2537 | -1.7199 |
| **V** | 1.53 | 4.6973 | 3.7786 | -0.9187 | -8.1458 | -1.7110 |
| **Mo** | 1.54 | 4.5840 | 3.6708 | -0.9131 | -8.0929 | -1.7067 |
| **Ti** | 1.60 | 3.9546 | 3.0738 | -0.8808 | -7.7894 | -1.6811 |
| **W** | 1.62 | 3.7626 | 2.8923 | -0.8703 | -7.6932 | -1.6727 |
| **Nb** | 1.64 | 3.5788 | 2.7189 | -0.8599 | -7.5994 | -1.6645 |
| **Na** | 1.66 | 3.4027 | 2.5530 | -0.8497 | -7.5079 | -1.6564 |
| **Cm** | 1.69 | 3.1521 | 2.3176 | -0.8345 | -7.3746 | -1.64434 |
| **Sc** | 1.70 | 3.0720 | 2.2425 | -0.8295 | -7.3312 | -1.6404 |
| **Ta** | 1.70 | 3.0720 | 2.2425 | -0.8295 | -7.3312 | -1.6404 |
| **Zr** | 1.75 | 2.6956 | 1.8906 | -0.8050 | -7.1217 | -1.6210 |
| **hf** | 1.75 | 2.6956 | 1.8906 | -0.8050 | -7.1217 | -1.6210 |
| **Ca** | 1.76 | 2.6248 | 1.8246 | -0.8002 | -7.0813 | -1.6172 |
| **Am** | 1.80 | 2.3557 | 1.5745 | -0.7812 | -6.9239 | -1.6022 |
| **Lu** | 1.87 | 1.9332 | 1.1843 | -0.7489 | -6.6647 | -1.5767 |
| Yb | 1.87 | 1.9332 | 1.1843 | -0.7489 | -6.6647 | -1.5767 |
| Pu | 1.87 | 1.9332 | 1.1843 | -0.7489 | -6.6647 | -1.5767 |
| Er | 1.89 | 1.8228 | 1.0829 | -0.7399 | -6.5942 | -1.5696 |
| Y | 1.90 | 1.7691 | 1.0337 | -0.7354 | -6.5595 | -1.5661 |
| Tm | 1.90 | 1.7691 | 1.0337 | -0.7354 | -6.5595 | -1.5661 |
| Np | 1.90 | 1.7691 | 1.0337 | -0.7354 | -6.5595 | -1.5661 |
| Dy | 1.92 | 1.6649 | 0.9383 | -0.7266 | -6.4912 | -1.5591 |
| Ho | 1.92 | 1.6649 | 0.9383 | -0.7266 | -6.4912 | -1.5591 |
| Tb | 1.94 | 1.5645 | 0.8467 | -0.7178 | -6.4242 | -1.5522 |
| Sr | 1.95 | 1.5158 | 0.8023 | -0.7135 | -6.3913 | -1.5488 |
| Gd | 1.96 | 1.4679 | 0.7588 | -0.7091 | -6.3587 | -1.5454 |
| U | 1.96 | 1.4679 | 0.7588 | -0.7091 | -6.3587 | -1.5454 |
| Sm | 1.98 | 1.3748 | 0.6743 | -0.7006 | -6.2945 | -1.5386 |
| Eu | 1.98 | 1.3748 | 0.6743 | -0.7006 | -6.2945 | -1.5386 |
| Pm | 1.99 | 1.3296 | 0.6333 | -0.6963 | -6.2628 | -1.5353 |
| Pa | 2.00 | 1.2852 | 0.5931 | -0.6921 | -6.2315 | -1.5319 |
| Nd | 2.01 | 1.2416 | 0.5538 | -0.6878 | -6.2005 | -1.5286 |
| K | 2.03 | 1.1568 | 0.4773 | -0.6795 | -6.1394 | -1.5220 |
| Pr | 2.03 | 1.1568 | 0.4773 | -0.6795 | -6.1394 | -1.5220 |
| Ce | 2.04 | 1.1156 | 0.4403 | -0.6753 | -6.1093 | -1.5188 |
| Th | 2.06 | 1.0353 | 0.3682 | -0.6670 | -6.0500 | -1.5123 |
| La | 2.07 | 0.9962 | 0.3333 | -0.6630 | -6.0208 | -1.5091 |
| Ba | 2.15 | 0.7079 | 0.0770 | -0.6309 | -5.7968 | -1.4840 |
| Ac | 2.15 | 0.7079 | 0.0770 | -0.6309 | -5.7968 | -1.4840 |
| Rb | 2.20 | 0.5474 | -0.0641 | -0.6114 | -5.6650 | -1.4688 |
| Ra | 2.21 | 0.5169 | -0.0907 | -0.6076 | -5.6394 | -1.4658 |
| Cs | 2.44 | -0.0572 | -0.5810 | -0.5238 | -5.1078 | -1.4008 |
| Fr | 2.60 | -0.3452 | -0.8153 | -0.4701 | -4.7935 | -1.3595 |

**Table 2** Ground-state energies from Hartree–Fock approximation (HFA) and random phase approximation (RPA) of nonmetallic elements. $\varepsilon_x$ and $\varepsilon_c$ represent the exchange and correlation energies.

| elements | $r_s$ (Å) | HFA | RPA | PRA -HFA | $\varepsilon_x$ (eV) | $\varepsilon_c$ (eV) |
|---|---|---|---|---|---|---|
| **He** | 0.28 | 338.8759 | 336.5207 | -2.3552 | -44.5109 | -2.4902 |
| **H** | 0.31 | 272.5717 | 270.3026 | -2.2691 | -40.2034 | -2.4139 |
| **F** | 0.57 | 70.6530 | 68.8991 | -1.7539 | -21.8650 | -1.9784 |
| **Ne** | 0.58 | 67.8674 | 66.1282 | -1.7392 | -21.4880 | -1.9666 |
| **O** | 0.66 | 50.1239 | 48.4940 | -1.6299 | -18.8834 | -1.8803 |
| **N** | 0.71 | 42.0772 | 40.5091 | -1.5681 | -17.5536 | -1.8326 |
| **C** | 0.76 | 35.6444 | 34.1339 | -1.5106 | -16.3987 | -1.7889 |
| **B** | 0.84 | 27.7659 | 26.3400 | -1.4259 | -14.8370 | -1.7261 |
| **Cl** | 1.02 | 16.6756 | 15.4139 | -1.2616 | -12.2187 | -1.9808 |
| **S** | 1.05 | 15.3973 | 14.1602 | -1.2371 | -11.8696 | -1.9617 |
| **Ar** | 1.06 | 14.9973 | 13.7682 | -1.2291 | -11.7576 | -1.9555 |
| **P** | 1.07 | 14.6095 | 13.3883 | -1.2212 | -11.6477 | -1.9493 |
| **Si** | 1.11 | 13.1711 | 11.9809 | -1.1901 | -11.2280 | -1.9251 |
| **Kr** | 1.16 | 11.5972 | 10.4443 | -1.1528 | -10.7440 | -1.8960 |
| **As** | 1.19 | 10.7559 | 9.6247 | -1.1312 | -10.4731 | -1.8790 |
| **Br** | 1.20 | 10.4909 | 9.3667 | -1.1242 | -10.3859 | -1.8734 |
| **Se** | 1.20 | 10.4909 | 9.3667 | -1.1242 | -10.3859 | -1.8734 |
| **Te** | 1.38 | 6.7552 | 5.7492 | -1.0059 | -9.0312 | -1.7802 |
| **I** | 1.39 | 6.5938 | 5.5940 | -0.9998 | -8.9662 | -1.7753 |
| **Xe** | 1.40 | 6.4364 | 5.4426 | -0.9938 | -8.9022 | -1.7705 |
| **At** | 1.50 | 5.0532 | 4.1177 | -0.9354 | -8.3087 | -1.7243 |
| **Rn** | 1.50 | 5.0532 | 4.1177 | 0.9354 | -8.3087 | -1.7243 |

**Table 3** Thomas–Fermi calculation of the total potential energy, where $\lambda$ represents

the shielding parameter.

| element | $r_s$ | $V_{TF}$ (eV) ($\lambda=0.1$) | $V_{TF}$ (eV) ($\lambda=0.3$) | $V_{TF}$ (eV) ($\lambda=0.5$) | $V_{TF}$ (eV) ($\lambda=0.7$) | $V_{TF}$ (eV) ($\lambda=1$) |
|---|---|---|---|---|---|---|
| He | 0.28 | 49.9393 | 47.2196 | 44.6480 | 42.2164 | 38.8151 |
| H | 0.31 | 44.9714 | 42.2678 | 39.7268 | 37.3385 | 34.0226 |
| F | 0.57 | 23.8304 | 21.2629 | 18.9720 | 16.9279 | 14.2672 |
| Ne | 0.58 | 23.3961 | 20.8337 | 18.5519 | 16.5200 | 13.8817 |
| O | 0.66 | 20.3964 | 17.8742 | 15.6639 | 13.7269 | 11.2611 |
| N | 0.71 | 18.8655 | 16.3681 | 14.2013 | 12.3213 | 9.9576 |
| C | 0.76 | 17.5364 | 15.0636 | 12.9394 | 11.1148 | 8.8488 |
| B | 0.84 | 15.7399 | 13.3057 | 11.2481 | 9.5086 | 7.3905 |
| Be | 0.96 | 13.6081 | 11.2309 | 9.2689 | 7.6497 | 5.7354 |
| Cl | 1.02 | 12.7310 | 10.3817 | 8.4658 | 6.9036 | 5.0837 |
| S | 1.05 | 12.3302 | 9.9947 | 8.1015 | 6.5670 | 4.7925 |
| Ar | 1.06 | 12.2017 | 9.8707 | 7.9851 | 6.4596 | 4.7001 |
| P | 1.07 | 12.0756 | 9.7492 | 7.8710 | 6.3546 | 4.6098 |
| Si | 1.11 | 11.5939 | 9.2858 | 7.4371 | 5.9565 | 4.2694 |
| Kr | 1.16 | 11.0389 | 8.7532 | 6.9408 | 5.5037 | 3.8862 |
| As | 1.19 | 10.7283 | 8.4561 | 6.6651 | 5.2535 | 3.6762 |
| Ge | 1.20 | 10.6283 | 8.3605 | 6.5766 | 5.1734 | 3.6093 |
| Br | 1.20 | 10.6283 | 8.3605 | 6.5766 | 5.1734 | 3.6093 |
| Se | 1.20 | 10.6283 | 8.3605 | 6.5766 | 5.1734 | 3.6093 |
| Al | 1.21 | 10.5299 | 8.2666 | 6.4897 | 5.0948 | 3.5439 |
| Zn | 1.22 | 10.4332 | 8.1743 | 6.4044 | 5.0178 | 3.4799 |
| Ga | 1.22 | 10.4332 | 8.1743 | 6.4044 | 5.0178 | 3.4799 |
| Ni | 1.24 | 10.2444 | 7.9943 | 6.2384 | 4.8682 | 3.3559 |
| Co | 1.26 | 10.0616 | 7.8204 | 6.0783 | 4.7244 | 3.2373 |
| Li | 1.28 | 9.8846 | 7.6521 | 5.9238 | 4.5859 | 3.1236 |

| element | $r_s$ | $V_{TF}$ (eV) | $V_{TF}$ (eV) | $V_{TF}$ (eV) | $V_{TF}$ (eV) | $V_{TF}$ (eV) |
| **Fe** | 1.32 | 9.5468 | 7.3317 | 5.6306 | 4.3241 | 2.9102 |
| **Cu** | 1.32 | 9.5468 | 7.3317 | 5.6306 | 4.3241 | 2.9102 |
| **Hg** | 1.32 | 9.5468 | 7.3317 | 5.6306 | 4.3241 | 2.9102 |
| **Pt** | 1.36 | 9.2291 | 7.0312 | 5.3568 | 4.0811 | 2.7138 |
| **Au** | 1.36 | 9.2291 | 7.0312 | 5.3568 | 4.0811 | 2.7138 |
| **Te** | 1.38 | 9.0771 | 6.8879 | 5.2266 | 3.9660 | 2.6215 |
| **Cr** | 1.39 | 9.0028 | 6.8178 | 5.1631 | 3.9100 | 2.5768 |
| **Mn** | 1.39 | 9.0028 | 6.8178 | 5.1631 | 3.9100 | 2.5768 |
| **Pd** | 1.39 | 9.0028 | 6.8178 | 5.1631 | 3.9100 | 2.5768 |
| **Sn** | 1.39 | 9.0028 | 6.8178 | 5.1631 | 3.9100 | 2.5768 |
| **Sb** | 1.39 | 9.0028 | 6.8178 | 5.1631 | 3.9100 | 2.5768 |
| **I** | 1.39 | 9.0028 | 6.8178 | 5.1631 | 3.9100 | 2.5768 |
| **Po** | 1.40 | 8.9296 | 6.7488 | 5.1007 | 3.8550 | 2.5329 |
| **Xe** | 1.40 | 8.9296 | 6.7488 | 5.1007 | 3.8550 | 2.5329 |
| **Mg** | 1.41 | 8.8574 | 6.6809 | 5.0392 | 3.8010 | 2.4899 |
| **Ir** | 1.41 | 8.8574 | 6.6809 | 5.0392 | 3.8010 | 2.4899 |
| **Rh** | 1.42 | 8.7862 | 6.6140 | 4.9788 | 3.7479 | 2.4478 |
| **In** | 1.42 | 8.7862 | 6.6140 | 4.9788 | 3.7479 | 2.4478 |
| **Cd** | 1.44 | 8.6469 | 6.4831 | 4.8608 | 3.6444 | 2.3660 |
| **Os** | 1.44 | 8.6469 | 6.4831 | 4.8608 | 3.6444 | 2.3660 |
| **Ag** | 1.45 | 8.5787 | 6.4191 | 4.8032 | 3.5940 | 2.3263 |
| **Ti** | 1.45 | 8.5787 | 6.4191 | 4.8032 | 3.5940 | 2.3263 |
| **Ru** | 1.46 | 8.5114 | 6.3560 | 4.7465 | 3.5445 | 2.2874 |
| **Pb** | 1.46 | 8.5114 | 6.3560 | 4.7465 | 3.5445 | 2.2874 |
| **Tc** | 1.47 | 8.4450 | 6.2939 | 4.6907 | 3.4959 | 2.2492 |
| **Bi** | 1.48 | 8.3796 | 6.2326 | 4.6358 | 3.4480 | 2.2118 |
| **At** | 1.50 | 8.2514 | 6.1128 | 4.5284 | 3.3548 | 2.1391 |
| **Rn** | 1.50 | 8.2514 | 6.1128 | 4.5284 | 3.3548 | 2.1391 |
| **Re** | 1.51 | 8.1885 | 6.0541 | 4.4760 | 3.3093 | 2.1038 |
| **V** | 1.53 | 8.0653 | 5.9392 | 4.3735 | 3.2206 | 2.0352 |
| **Mo** | 1.54 | 8.0050 | 5.8830 | 4.3235 | 3.1774 | 2.0018 |
| **Ti** | 1.60 | 7.6587 | 5.5613 | 4.0384 | 2.9325 | 1.8146 |
| **W** | 1.62 | 7.5490 | 5.4598 | 3.9488 | 2.8560 | 1.7567 |
| **Nb** | 1.64 | 7.4421 | 5.3610 | 3.8619 | 2.7819 | 1.7009 |
| **Na** | 1.66 | 7.3377 | 5.2647 | 3.7774 | 2.7102 | 1.6471 |
| **Cm** | 1.69 | 7.1859 | 5.1249 | 3.6551 | 2.6068 | 1.5701 |
| **Sc** | 1.70 | 7.1364 | 5.0795 | 3.6154 | 2.5734 | 1.5453 |
| **Ta** | 1.70 | 7.1364 | 5.0795 | 3.6154 | 2.5734 | 1.5453 |
| **Zr** | 1.75 | 6.8980 | 4.8609 | 3.4254 | 2.4139 | 1.4279 |
| **hf** | 1.75 | 6.8980 | 4.8609 | 3.4254 | 2.4139 | 1.4279 |
| **Ca** | 1.76 | 6.8519 | 4.8188 | 3.3890 | 2.3834 | 1.4057 |
| **Am** | 1.80 | 6.6729 | 4.6555 | 3.2481 | 2.2661 | 1.3206 |
| **Lu** | 1.87 | 6.3783 | 4.3881 | 3.0189 | 2.0770 | 1.1852 |
| **Yb** | 1.87 | 6.3783 | 4.3881 | 3.0189 | 2.0770 | 1.1852 |
| **Pu** | 1.87 | 6.3783 | 4.3881 | 3.0189 | 2.0770 | 1.1852 |
| **Er** | 1.89 | 6.2982 | 4.3157 | 2.9573 | 2.0264 | 1.1494 |
| **Y** | 1.90 | 6.2588 | 4.2802 | 2.9270 | 2.0017 | 1.1320 |
| **Tm** | 1.90 | 6.2588 | 4.2802 | 2.9270 | 2.0017 | 1.1320 |
| **Np** | 1.90 | 6.2588 | 4.2802 | 2.9270 | 2.0017 | 1.1320 |
| **Dy** | 1.92 | 6.1812 | 4.2102 | 2.8677 | 1.9533 | 1.0980 |
| **Ho** | 1.92 | 6.1812 | 4.2102 | 2.8677 | 1.9533 | 1.0980 |
| **Tb** | 1.94 | 6.1053 | 4.1419 | 2.8099 | 1.9063 | 1.0652 |
| **Sr** | 1.95 | 6.0679 | 4.1083 | 2.7816 | 1.8833 | 1.0492 |
| **Gd** | 1.96 | 6.0309 | 4.0751 | 2.7536 | 1.8606 | 1.0334 |
| **U** | 1.96 | 6.0309 | 4.0751 | 2.7536 | 1.8606 | 1.0334 |
| **Sm** | 1.98 | 5.9581 | 4.0098 | 2.6986 | 1.8162 | 1.0027 |
| **Eu** | 1.98 | 5.9581 | 4.0098 | 2.6986 | 1.8162 | 1.0027 |
| **Pm** | 1.99 | 5.9222 | 3.9777 | 2.6717 | 1.7945 | 0.9878 |

| Pa | 2.00 | 5.8867 | 3.9460 | 2.6451 | 1.7730 | 0.9731 |
| --- | --- | --- | --- | --- | --- | --- |
| Nd | 2.01 | 5.8516 | 3.9146 | 2.6188 | 1.7519 | 0.9586 |
| K | 2.03 | 5.7823 | 3.8528 | 2.5672 | 1.7105 | 0.9304 |
| Pr | 2.03 | 5.7823 | 3.8528 | 2.5672 | 1.7105 | 0.9304 |
| Ce | 2.04 | 5.7482 | 3.8225 | 2.5419 | 1.6903 | 0.9166 |
| Th | 2.06 | 5.6811 | 3.7627 | 2.4921 | 1.6506 | 0.8897 |
| La | 2.07 | 5.6480 | 3.7333 | 2.4677 | 1.6312 | 0.8766 |
| Ba | 2.15 | 5.3945 | 3.5092 | 2.2827 | 1.4849 | 0.7791 |
| Ac | 2.15 | 5.3945 | 3.5092 | 2.2827 | 1.4849 | 0.7791 |
| Rb | 2.20 | 5.2456 | 3.3783 | 2.1758 | 1.4013 | 0.7243 |
| Ra | 2.21 | 5.2166 | 3.3530 | 2.1551 | 1.3852 | 0.7138 |
| Cs | 2.44 | 4.6175 | 2.8344 | 1.7399 | 1.0681 | 0.5137 |
| Fr | 2.60 | 4.2645 | 2.5353 | 1.5073 | 0.8961 | 0.4108 |

**Table 4** Wigner correlation potential $V_c(eV)$.

| element | $r_s(\text{Å})$ | $V_c(eV)$ | $V_c(eV)$ ($\lambda=0.1$) | $V_c(eV)$ ($\lambda=0.3$) | $V_c(eV)$ ($\lambda=0.5$) | $V_c(eV)$ ($\lambda=0.7$) | $V_c(eV)$ ($\lambda=1$) |
| --- | --- | --- | --- | --- | --- | --- | --- |
| He | 0.28 | -1.8691 | -1.8175 | -1.7185 | -1.6249 | -1.5364 | -1.4126 |
| H | 0.31 | -1.8133 | -1.7579 | -1.6522 | -1.5529 | -1.4596 | -1.3299 |
| F | 0.57 | -1.5343 | -1.4492 | -1.2931 | -1.1538 | -1.0295 | -0.8677 |
| Ne | 0.58 | -1.5271 | -1.4411 | -1.2833 | -1.1427 | -1.0176 | -0.8550 |
| O | 0.66 | -1.4750 | -1.3808 | -1.2101 | -1.0604 | -0.9293 | -0.7624 |
| N | 0.71 | -1.4459 | -1.3468 | -1.1685 | -1.0138 | -0.8796 | -0.7109 |
| C | 0.76 | -1.4190 | -1.3151 | -1.1297 | -0.9704 | -0.8335 | -0.6636 |
| B | 0.84 | -1.3793 | -1.2682 | -1.0721 | -0.9063 | -0.7661 | -0.5955 |
| Be | 0.96 | -1.3263 | -1.2049 | -0.9944 | -0.8207 | -0.6773 | -0.5078 |
| Cl | 1.02 | -1.3020 | -1.1757 | -0.9587 | -0.7818 | -0.6375 | -0.4695 |
| S | 1.05 | -1.2903 | -1.1617 | -0.9416 | -0.7633 | -0.6187 | -0.4515 |
| **Ar** | 1.06 | -1.2864 | -1.1571 | -0.9360 | -0.7572 | -0.6126 | -0.4457 |
| **P**  | 1.07 | -1.2826 | -1.1525 | -0.9305 | -0.7512 | -0.6065 | -0.4400 |
| **Si** | 1.11 | -1.2678 | -1.1346 | -0.9087 | -0.7278 | -0.5829 | -0.4178 |
| **Kr** | 1.16 | -1.2498 | -1.1129 | -0.8825 | -0.6997 | -0.5549 | -0.3918 |
| **As** | 1.19 | -1.2393 | -1.1002 | -0.8672 | -0.6835 | -0.5388 | -0.3770 |
| **Ge** | 1.20 | -1.2358 | -1.0961 | -0.8622 | -0.6782 | -0.5335 | -0.3722 |
| **Br** | 1.20 | -1.2358 | -1.0961 | -0.8622 | -0.6782 | -0.5335 | -0.3722 |
| **Se** | 1.20 | -1.2358 | -1.0961 | -0.8622 | -0.6782 | -0.5335 | -0.3722 |
| **Al** | 1.21 | -1.2324 | -1.0920 | -0.8572 | -0.6730 | -0.5283 | -0.3675 |
| **Zn** | 1.22 | -1.2290 | -1.0879 | -0.8523 | -0.6678 | -0.5232 | -0.3628 |
| **Ga** | 1.22 | -1.2290 | -1.0879 | -0.8523 | -0.6678 | -0.5232 | -0.3628 |
| **Ni** | 1.24 | -1.2223 | -1.0797 | -0.8426 | -0.6575 | -0.5131 | -0.3537 |
| **Co** | 1.26 | -1.2156 | -1.0717 | -0.8330 | -0.6474 | -0.5032 | -0.3448 |
| **Li** | 1.28 | -1.2091 | -1.0638 | -0.8235 | -0.6375 | -0.4935 | -0.3362 |
| **Fe** | 1.32 | -1.1962 | -1.0483 | -0.8051 | -0.6183 | -0.4748 | -0.3195 |
| **Cu** | 1.32 | -1.1962 | -1.0483 | -0.8051 | -0.6183 | -0.4748 | -0.3195 |
| **Hg** | 1.32 | -1.1962 | -1.0483 | -0.8051 | -0.6183 | -0.4748 | -0.3195 |
| **Pt** | 1.36 | -1.1836 | -1.0331 | -0.7871 | -0.5997 | -0.4568 | -0.3038 |
| **Au** | 1.36 | -1.1836 | -1.0331 | -0.7871 | -0.5997 | -0.4568 | -0.3038 |
| **Te** | 1.38 | -1.1775 | -1.0257 | -0.7783 | -0.5906 | -0.4481 | -0.2962 |
| **Cr** | 1.39 | -1.1744 | -1.0220 | -0.7740 | -0.5861 | -0.4439 | -0.2925 |
| **Mn** | 1.39 | -1.1744 | -1.0220 | -0.7740 | -0.5861 | -0.4439 | -0.2925 |
| **Pd** | 1.39 | -1.1744 | -1.0220 | -0.7740 | -0.5861 | -0.4439 | -0.2925 |
| **Sn** | 1.39 | -1.1744 | -1.0220 | -0.7740 | -0.5861 | -0.4439 | -0.2925 |
| **Sb** | 1.39 | -1.1744 | -1.0220 | -0.7740 | -0.5861 | -0.4439 | -0.2925 |
| **I**  | 1.39 | -1.1744 | -1.0220 | -0.7740 | -0.5861 | -0.4439 | -0.2925 |
| **Po** | 1.40 | -1.1714 | -1.0183 | -0.7696 | -0.5817 | -0.4396 | -0.2889 |
| **Xe** | 1.40 | -1.1714 | -1.0183 | -0.7696 | -0.5817 | -0.4396 | -0.2889 |
| **Mg** | 1.41 | -1.1684 | -1.0147 | -0.7654 | -0.5773 | -0.4354 | -0.2852 |
| **Ir** | 1.41 | -1.1684 | -1.0147 | -0.7654 | -0.5773 | -0.4354 | -0.2852 |
| **Rh** | 1.42 | -1.1653 | -1.0111 | -0.7611 | -0.5729 | -0.4313 | -0.2817 |
| **In** | 1.42 | -1.1653 | -1.0111 | -0.7611 | -0.5729 | -0.4313 | -0.2817 |
| **Cd** | 1.44 | -1.1594 | -1.0039 | -0.7527 | -0.5643 | -0.4231 | -0.2747 |
| **Os** | 1.44 | -1.1594 | -1.0039 | -0.7527 | -0.5643 | -0.4231 | -0.2747 |
| **Ag** | 1.45 | -1.1564 | -1.0003 | -0.7485 | -0.5601 | -0.4191 | -0.2713 |
| **Ti** | 1.45 | -1.1564 | -1.0003 | -0.7485 | -0.5601 | -0.4191 | -0.2713 |
| **Ru** | 1.46 | -1.1535 | -0.9968 | -0.7444 | -0.5559 | -0.4151 | -0.2679 |
| **Pb** | 1.46 | -1.1535 | -0.9968 | -0.7444 | -0.5559 | -0.4151 | -0.2679 |
| **Tc** | 1.47 | -1.1506 | -0.9933 | -0.7403 | -0.5517 | -0.4112 | -0.2645 |
| **Bi** | 1.48 | -1.1477 | -0.9898 | -0.7362 | -0.5476 | -0.4073 | -0.2613 |
| **At** | 1.50 | -1.1419 | -0.9828 | -0.7281 | -0.5394 | -0.3996 | -0.2548 |
| **Rn** | 1.50 | -1.1419 | -0.9828 | -0.7281 | -0.5394 | -0.3996 | -0.2548 |
| **Re** | 1.51 | -1.1390 | -0.9794 | -0.7241 | -0.5354 | -0.3958 | -0.2516 |
| **V** | 1.53 | -1.1334 | -0.9726 | -0.7162 | -0.5274 | -0.3884 | -0.2454 |
| **Mo** | 1.54 | -1.1305 | -0.9692 | -0.7123 | -0.5235 | -0.3847 | -0.2424 |
| **Ti** | 1.60 | -1.1139 | -0.9492 | -0.6893 | -0.5005 | -0.3635 | -0.2249 |
| **W** | 1.62 | -1.1085 | -0.9427 | -0.6818 | -0.4931 | -0.3567 | -0.2194 |
| **Nb** | 1.64 | -1.1031 | -0.9363 | -0.6745 | -0.4859 | -0.3500 | -0.2140 |
| **Na** | 1.66 | -1.0978 | -0.9299 | -0.6672 | -0.4787 | -0.3435 | -0.2087 |
| **Cm** | 1.69 | -1.0899 | -0.9204 | -0.6565 | -0.4682 | -0.3339 | -0.2011 |
| **Sc** | 1.70 | -1.0873 | -0.9173 | -0.6529 | -0.4647 | -0.3308 | -0.1986 |
| **Ta** | 1.70 | -1.0873 | -0.9173 | -0.6529 | -0.4647 | -0.3308 | -0.1986 |
| **Zr** | 1.75 | -1.0745 | -0.9020 | -0.6356 | -0.4479 | -0.3156 | -0.1867 |
| **hf** | 1.75 | -1.0745 | -0.9020 | -0.6356 | -0.4479 | -0.3156 | -0.1867 |
| **Ca** | 1.76 | -1.0719 | -0.8990 | -0.6322 | -0.4446 | -0.3127 | -0.1844 |
| **Am** | 1.80 | -1.0619 | -0.8870 | -0.6188 | -0.4317 | -0.3012 | -0.1755 |
| **Lu** | 1.87 | -1.0448 | -0.8666 | -0.5962 | -0.4102 | -0.2822 | -0.1610 |
| **Yb** | 1.87 | -1.0448 | -0.8666 | -0.5962 | -0.4102 | -0.2822 | -0.1610 |
| **Pu** | 1.87 | -1.0448 | -0.8666 | -0.5962 | -0.4102 | -0.2822 | -0.1610 |
| **Er** | 1.89 | -1.0400 | -0.8609 | -0.5899 | -0.4042 | -0.2770 | -0.1571 |
| **Y** | 1.90 | -1.0376 | -0.8581 | -0.5868 | -0.4013 | -0.2744 | -0.1552 |
| **Tm** | 1.90 | -1.0376 | -0.8581 | -0.5868 | -0.4013 | -0.2744 | -0.1552 |
| **Np** | 1.90 | -1.0376 | -0.8581 | -0.5868 | -0.4013 | -0.2744 | -0.1552 |
| **Dy** | 1.92 | -1.0329 | -0.8525 | -0.5806 | -0.3955 | -0.2694 | -0.1514 |
| **Ho** | 1.92 | -1.0329 | -0.8525 | -0.5806 | -0.3955 | -0.2694 | -0.1514 |
| **Tb** | 1.94 | -1.0282 | -0.8469 | -0.5745 | -0.3898 | -0.2644 | -0.1478 |
| **Sr** | 1.95 | -1.0259 | -0.8441 | -0.5715 | -0.3870 | -0.2620 | -0.1460 |
| **Gd** | 1.96 | -1.0236 | -0.8414 | -0.5685 | -0.3842 | -0.2596 | -0.1442 |
| **U** | 1.96 | -1.0236 | -0.8414 | -0.5685 | -0.3842 | -0.2596 | -0.1442 |
| **Sm** | 1.98 | -1.0189 | -0.8359 | -0.5626 | -0.3786 | -0.2548 | -0.1407 |
| **Eu** | 1.98 | -1.0189 | -0.8359 | -0.5626 | -0.3786 | -0.2548 | -0.1407 |
| **Pm** | 1.99 | -1.0166 | -0.8332 | -0.5596 | -0.3759 | -0.2525 | -0.1390 |
| **Pa** | 2.00 | -1.0144 | -0.8305 | -0.5567 | -0.3732 | -0.2501 | -0.1373 |
| **Nd** | 2.01 | -1.0121 | -0.8278 | -0.5538 | -0.3705 | -0.2478 | -0.1356 |
| **K** | 2.03 | -1.0076 | -0.8224 | -0.5480 | -0.3651 | -0.2433 | -0.1323 |
| **Pr** | 2.03 | -1.0076 | -0.8224 | -0.5480 | -0.3651 | -0.2433 | -0.1323 |
| **Ce** | 2.04 | -1.0053 | -0.8198 | -0.5451 | -0.3625 | -0.2411 | -0.1307 |
| **Th** | 2.06 | -1.0008 | -0.8145 | -0.5395 | -0.3573 | -0.2367 | -0.1276 |
| **La** | 2.07 | -0.9986 | -0.8119 | -0.5367 | -0.3547 | -0.2345 | -0.1260 |
| **Ba** | 2.15 | -0.9812 | -0.7913 | -0.5148 | -0.3349 | -0.2178 | -0.1143 |
| **Ac** | 2.15 | -0.9812 | -0.7913 | -0.5148 | -0.3349 | -0.2178 | -0.1143 |
| **Rb** | 2.20 | -0.9705 | -0.7789 | -0.5016 | -0.3231 | -0.2081 | -0.1075 |
| **Ra** | 2.21 | -0.9684 | -0.7764 | -0.4990 | -0.3207 | -0.2062 | -0.1062 |
| **Cs** | 2.44 | -0.9221 | -0.7225 | -0.4435 | -0.2722 | -0.1671 | -0.0804 |
| **Fr** | 2.60 | -0.8921 | -0.6878 | -0.4089 | -0.2431 | -0.1445 | -0.0663 |

**Table 5** Hedin–Lundqvist correlation potential $V_c(eV)$.

| element | $r_s(Å)$ | $V_c(eV)$ | $V_c(eV)$ ($\lambda=0.1$) | $V_c(eV)$ ($\lambda=0.3$) | $V_c(eV)$ ($\lambda=0.5$) | $V_c(eV)$ ($\lambda=0.7$) | $V_c(eV)$ ($\lambda=1$) |
|---|---|---|---|---|---|---|---|
| He | 0.28 | -1.3252 | -1.2886 | -1.2184 | -1.1521 | -1.0893 | -1.0016 |
| H  | 0.31 | -1.2945 | -1.2550 | -1.1795 | -1.1086 | -1.0420 | -0.9494 |
| F  | 0.57 | -1.1118 | -1.0502 | -0.9371 | -0.8361 | -0.7460 | -0.6288 |
| Ne | 0.58 | -1.1066 | -1.0443 | -0.9299 | -0.8281 | -0.7374 | -0.6196 |
| O  | 0.66 | -1.0682 | -1.0000 | -0.8764 | -0.7680 | -0.6730 | -0.5521 |
| N  | 0.71 | -1.0466 | -0.9749 | -0.8458 | -0.7338 | -0.6367 | -0.5146 |
| C  | 0.76 | -1.0265 | -0.9514 | -0.8172 | -0.7020 | -0.6030 | -0.4801 |
| B  | 0.84 | -0.9970 | -0.9167 | -0.7749 | -0.6551 | -0.5538 | -0.4304 |
| Be | 0.96 | -0.9578 | -0.8701 | -0.7181 | -0.5927 | -0.4891 | -0.3667 |
| Cl | 1.02 | -0.9401 | -0.8489 | -0.6923 | -0.5645 | -0.4603 | -0.3390 |
| S  | 1.05 | -0.9316 | -0.8388 | -0.6799 | -0.5511 | -0.4467 | -0.3260 |
| Ar | 1.06 | -0.9289 | -0.8354 | -0.6758 | -0.5467 | -0.4423 | -0.3218 |
| P  | 1.07 | -0.9261 | -0.8321 | -0.6718 | -0.5424 | -0.4379 | -0.3177 |
| Si | 1.11 | -0.9155 | -0.8193 | -0.6562 | -0.5255 | -0.4209 | -0.3017 |
| Kr | 1.16 | -0.9027 | -0.8038 | -0.6374 | -0.5054 | -0.4008 | -0.2830 |
| As | 1.19 | -0.8953 | -0.7948 | -0.6265 | -0.4938 | -0.3892 | -0.2724 |
| Ge | 1.20 | -0.8928 | -0.7919 | -0.6229 | -0.4900 | -0.3854 | -0.2689 |
| Br | 1.20 | -0.8928 | -0.7919 | -0.6229 | -0.4900 | -0.3854 | -0.2689 |
| Se | 1.20 | -0.8928 | -0.7919 | -0.6229 | -0.4900 | -0.3854 | -0.2689 |
| Al | 1.21 | -0.8904 | -0.7890 | -0.6194 | -0.4862 | -0.3817 | -0.2655 |
| Zn | 1.22 | -0.8881 | -0.7861 | -0.6159 | -0.4825 | -0.3781 | -0.2622 |
| Ga | 1.22 | -0.8881 | -0.7861 | -0.6159 | -0.4825 | -0.3781 | -0.2622 |
| Ni | 1.24 | -0.8834 | -0.7803 | -0.6089 | -0.4752 | -0.3708 | -0.2556 |
| Co | 1.26 | -0.8787 | -0.7747 | -0.6021 | -0.4680 | -0.3638 | -0.2493 |
| **Li** | 1.28 | -0.8742 | -0.7692 | -0.5954 | -0.4610 | -0.3568 | -0.2431 |
| **Fe** | 1.32 | -0.8653 | -0.7583 | -0.5824 | -0.4472 | -0.3435 | -0.2312 |
| **Cu** | 1.32 | -0.8653 | -0.7583 | -0.5824 | -0.4472 | -0.3435 | -0.2312 |
| **Hg** | 1.32 | -0.8653 | -0.7583 | -0.5824 | -0.4472 | -0.3435 | -0.2312 |
| **Pt** | 1.36 | -0.8567 | -0.7478 | -0.5697 | -0.4340 | -0.3307 | -0.2199 |
| **Au** | 1.36 | -0.8567 | -0.7478 | -0.5697 | -0.4340 | -0.3307 | -0.2199 |
| **Te** | 1.38 | -0.8525 | -0.7426 | -0.5635 | -0.4276 | -0.3245 | -0.2145 |
| **Cr** | 1.39 | -0.8505 | -0.7401 | -0.5605 | -0.4244 | -0.3214 | -0.2118 |
| **Mn** | 1.39 | -0.8505 | -0.7401 | -0.5605 | -0.4244 | -0.3214 | -0.2118 |
| **Pd** | 1.39 | -0.8505 | -0.7401 | -0.5605 | -0.4244 | -0.3214 | -0.2118 |
| **Sn** | 1.39 | -0.8505 | -0.7401 | -0.5605 | -0.4244 | -0.3214 | -0.2118 |
| **Sb** | 1.39 | -0.8505 | -0.7401 | -0.5605 | -0.4244 | -0.3214 | -0.2118 |
| **I** | 1.39 | -0.8505 | -0.7401 | -0.5605 | -0.4244 | -0.3214 | -0.2118 |
| **Po** | 1.40 | -0.8484 | -0.7376 | -0.5574 | -0.4213 | -0.3184 | -0.2092 |
| **Xe** | 1.40 | -0.8484 | -0.7376 | -0.5574 | -0.4213 | -0.3184 | -0.2092 |
| **Mg** | 1.41 | -0.8464 | -0.7351 | -0.5544 | -0.4182 | -0.3154 | -0.2066 |
| **Ir** | 1.41 | -0.8464 | -0.7351 | -0.5544 | -0.4182 | -0.3154 | -0.2066 |
| **Rh** | 1.42 | -0.8443 | -0.7326 | -0.5515 | -0.4151 | -0.3125 | -0.2041 |
| **In** | 1.42 | -0.8443 | -0.7326 | -0.5515 | -0.4151 | -0.3125 | -0.2041 |
| **Cd** | 1.44 | -0.8403 | -0.7276 | -0.5456 | -0.4090 | -0.3067 | -0.1991 |
| **Os** | 1.44 | -0.8403 | -0.7276 | -0.5456 | -0.4090 | -0.3067 | -0.1991 |
| **Ag** | 1.45 | -0.8384 | -0.7252 | -0.5426 | -0.4060 | -0.3038 | -0.1967 |
| **Ti** | 1.45 | -0.8384 | -0.7252 | -0.5426 | -0.4060 | -0.3038 | -0.1967 |
| **Ru** | 1.46 | -0.8364 | -0.7228 | -0.5397 | -0.4031 | -0.3010 | -0.1942 |
| **Pb** | 1.46 | -0.8364 | -0.7228 | -0.5397 | -0.4031 | -0.3010 | -0.1942 |
| **Tc** | 1.47 | -0.8344 | -0.7204 | -0.5369 | -0.4001 | -0.2982 | -0.1919 |
| **Bi** | 1.48 | -0.8325 | -0.7180 | -0.5340 | -0.3972 | -0.2954 | -0.1895 |
| **At** | 1.50 | -0.8287 | -0.7132 | -0.5284 | -0.3914 | -0.2900 | -0.1849 |
| **Rn** | 1.50 | -0.8287 | -0.7132 | -0.5284 | -0.3914 | -0.2900 | -0.1849 |
| **Re** | 1.51 | -0.8268 | -0.7109 | -0.5256 | -0.3886 | -0.2873 | -0.1826 |
| **V** | 1.53 | -0.8230 | -0.7063 | -0.5201 | -0.3830 | -0.2820 | -0.1782 |
| **Mo** | 1.54 | -0.8212 | -0.7040 | -0.5173 | -0.3802 | -0.2794 | -0.1760 |
| **Ti** | 1.60 | -0.8103 | -0.6905 | -0.5014 | -0.3641 | -0.2644 | -0.1636 |
| **W** | 1.62 | -0.8067 | -0.6861 | -0.4962 | -0.3589 | -0.2596 | -0.1597 |
| **Nb** | 1.64 | -0.8033 | -0.6818 | -0.4911 | -0.3538 | -0.2548 | -0.1558 |
| **Na** | 1.66 | -0.7998 | -0.6775 | -0.4861 | -0.3488 | -0.2502 | -0.1521 |
| **Cm** | 1.69 | -0.7947 | -0.6712 | -0.4787 | -0.3414 | -0.2435 | -0.1466 |
| **Sc** | 1.70 | -0.7931 | -0.6691 | -0.4762 | -0.3390 | -0.2413 | -0.1449 |
| **Ta** | 1.70 | -0.7931 | -0.6691 | -0.4762 | -0.3390 | -0.2413 | -0.1449 |
| **Zr** | 1.75 | -0.7849 | -0.6589 | -0.4643 | -0.3272 | -0.2306 | -0.1364 |
| **hf** | 1.75 | -0.7849 | -0.6589 | -0.4643 | -0.3272 | -0.2306 | -0.1364 |
| **Ca** | 1.76 | -0.7833 | -0.6569 | -0.4620 | -0.3249 | -0.2285 | -0.1348 |
| **Am** | 1.80 | -0.7769 | -0.6489 | -0.4528 | -0.3159 | -0.2204 | -0.1284 |
| **Lu** | 1.87 | -0.7662 | -0.6355 | -0.4372 | -0.3008 | -0.2069 | -0.1181 |
| **Yb** | 1.87 | -0.7662 | -0.6355 | -0.4372 | -0.3008 | -0.2069 | -0.1181 |
| **Pu** | 1.87 | -0.7662 | -0.6355 | -0.4372 | -0.3008 | -0.2069 | -0.1181 |
| **Er** | 1.89 | -0.7632 | -0.6318 | -0.4329 | -0.2966 | -0.2033 | -0.1153 |
| **Y** | 1.90 | -0.7617 | -0.6299 | -0.4308 | -0.2946 | -0.2015 | -0.1139 |
| **Tm** | 1.90 | -0.7617 | -0.6299 | -0.4308 | -0.2946 | -0.2015 | -0.1139 |
| **Np** | 1.90 | -0.7617 | -0.6299 | -0.4308 | -0.2946 | -0.2015 | -0.1139 |
| **Dy** | 1.92 | -0.7588 | -0.6262 | -0.4265 | -0.2905 | -0.1979 | -0.1112 |
| **Ho** | 1.92 | -0.7588 | -0.6262 | -0.4265 | -0.2905 | -0.1979 | -0.1112 |
| **Tb** | 1.94 | -0.7559 | -0.6226 | -0.4224 | -0.2865 | -0.1944 | -0.1086 |
| **Sr** | 1.95 | -0.7544 | -0.6208 | -0.4203 | -0.2846 | -0.1927 | -0.1073 |
| **Gd** | 1.96 | -0.7530 | -0.6190 | -0.4182 | -0.2826 | -0.1910 | -0.1061 |
| **U** | 1.96 | -0.7530 | -0.6190 | -0.4182 | -0.2826 | -0.1910 | -0.1061 |
| **Sm** | 1.98 | -0.7502 | -0.6154 | -0.4142 | -0.2787 | -0.1876 | -0.1036 |
| **Eu** | 1.98 | -0.7502 | -0.6154 | -0.4142 | -0.2787 | -0.1876 | -0.1036 |
| **Pm** | 1.99 | -0.7488 | -0.6136 | -0.4122 | -0.2768 | -0.1859 | -0.1024 |
| **Pa** | 2.00 | -0.7474 | -0.6119 | -0.4102 | -0.2749 | -0.1843 | -0.1011 |
| **Nd** | 2.01 | -0.7460 | -0.6101 | -0.4082 | -0.2731 | -0.1827 | -0.1000 |
| **K** | 2.03 | -0.7432 | -0.6067 | -0.4042 | -0.2693 | -0.1795 | -0.0976 |
| **Pr** | 2.03 | -0.7432 | -0.6067 | -0.4042 | -0.2693 | -0.1795 | -0.0976 |
| **Ce** | 2.04 | -0.7418 | -0.6049 | -0.4023 | -0.2675 | -0.1779 | -0.0965 |
| **Th** | 2.06 | -0.7391 | -0.6015 | -0.3984 | -0.2639 | -0.1748 | -0.0942 |
| **La** | 2.07 | -0.7378 | -0.5998 | -0.3965 | -0.2621 | -0.1732 | -0.0931 |
| **Ba** | 2.15 | -0.7272 | -0.5865 | -0.3815 | -0.2482 | -0.1615 | -0.0847 |
| **Ac** | 2.15 | -0.7272 | -0.5865 | -0.3815 | -0.2482 | -0.1615 | -0.0847 |
| **Rb** | 2.20 | -0.7208 | -0.5785 | -0.3726 | -0.2399 | -0.1545 | -0.0799 |
| **Ra** | 2.21 | -0.7196 | -0.5769 | -0.3708 | -0.2383 | -0.1532 | -0.0789 |
| **Cs** | 2.44 | -0.6923 | -0.5424 | -0.3330 | -0.2044 | -0.1255 | -0.0603 |
| **Fr** | 2.60 | -0.6750 | -0.5204 | -0.3094 | -0.1839 | -0.1094 | -0.0501 |

**References**

[1] E. Cancès, G. Stoltz, M. Lewin, The electronic ground-state energy problem: a new reduced density matrix approach, J Chem Phys, 125 (2006) 64101.

[2] D. Lee, Ground state energy at unitarity, Physical Review C, 78 (2008) 024001.

[3] L. Hedin, B.I. Lundqvist, Explicit local exchange-correlation potentials, Journal of Physics C: Solid State Physics, 4 (1971) 2064.

[4] S. Huzinaga, Analytical Methods in Hartree-Fock Self-Consistent Field Theory, Physical Review, 122 (1961) 131-138.

[5] Q. Dong, T. Wang, R. Gan, C. Tong, R. Xu, M. Shao, C. Li, Z. Wei, Separators Based on the Dynamic Tip-Occupying Electrostatic Shield Effect for Dendrite-Free Lithium-Metal Batteries, Advanced Sustainable Systems, 6 (2022) 2100386.

[6] D. McClain, J. Wu, N. Tavan, J. Jiao, C.M. McCarter, R.F. Richards, S. Mesarovic, C.D. Richards, D.F. Bahr, Electrostatic Shielding in Patterned Carbon Nanotube Field Emission Arrays, The Journal of Physical Chemistry C, 111 (2007) 7514-7520.

[7] J. Zheng, Z. Huang, Y. Zeng, W. Liu, B. Wei, Z. Qi, Z. Wang, C. Xia, H. Liang, Electrostatic Shielding Regulation of Magnetron Sputtered Al-Based Alloy Protective Coatings Enables Highly Reversible Zinc Anodes, Nano Letters, 22 (2022) 1017-1023.

[8] Z. Sun, M. Xiao, S. Wang, D. Han, S. Song, G. Chen, Y. Meng, Electrostatic shield effect: an effective way to suppress dissolution of polysulfide anions in lithium–sulfur battery, Journal of Materials Chemistry A, 2 (2014) 15938-15944.

[9] F. Trani, D. Ninno, G. Cantele, G. Iadonisi, K. Hameeuw, E. Degoli, S. Ossicini, Screening in semiconductor nanocrystals: Ab initio results and Thomas-Fermi theory, Physical Review B, 73 (2006) 245430.

[10] J. Shin, V. Meunier, A.P. Baddorf, S.V. Kalinin, Nonlinear transport imaging by scanning impedance microscopy, Applied Physics Letters, 85 (2004) 4240-4242.

[11] K. Lee, M.I.B. Utama, S. Kahn, A. Samudrala, N. Leconte, B. Yang, S. Wang, K. Watanabe, T. Taniguchi, M.V.P. Altoé, G. Zhang, A. Weber-Bargioni, M. Crommie, P.D. Ashby, J. Jung, F. Wang, A. Zettl, Ultrahigh-resolution scanning microwave impedance microscopy of moiré lattices and superstructures, Science Advances, 6   eabd1919.

[12] N. Ishida, Local Impedance Measurement by Direct Detection of Oscillating Electrostatic Potential Using Kelvin Probe Force Microscopy, The Journal of Physical Chemistry C, 126 (2022) 17627-17634.

[13] S.V. Kalinin, D.A. Bonnell, Scanning impedance microscopy of electroactive interfaces, Applied Physics Letters, 78 (2001) 1306-1308.

[14] M. Gell-Mann, K.A. Brueckner, Correlation Energy of an Electron Gas at High Density, Physical Review, 106 (1957) 364-368.

[15] D.M. Ceperley, B.J. Alder, Ground State of the Electron Gas by a Stochastic Method, Physical Review Letters, 45 (1980) 566-569.

[16] R. Resta, Thomas-Fermi dielectric screening in semiconductors, Physical Review B, 16 (1977) 2717-2722.

[17] E. Wigner, On the Interaction of Electrons in Metals, Physical Review, 46 (1934) 1002-1011.

[18] J.P. Perdew, A. Zunger, Self-interaction correction to density-functional approximations for many-electron systems, Physical Review B, 23 (1981) 5048-5079.

[19] E.W. Brown, J.L. DuBois, M. Holzmann, D.M. Ceperley, Exchange-correlation energy for the three-dimensional homogeneous electron gas at arbitrary temperature, Physical Review B, 88 (2013) 081102.

[20] P. Bhattarai, A. Patra, C. Shahi, J.P. Perdew, How accurate are the parametrized correlation energies of the uniform electron gas?, Physical Review B, 97 (2018) 195128.

[21] L. Scalfi, T. Dufils, K.G. Reeves, B. Rotenberg, M. Salanne, A semiclassical Thomas–Fermi model to tune the metallicity of electrodes in molecular simulations, The Journal of Chemical Physics, 153 (2020) 174704.

[22] M. Zarenia, D. Neilson, B. Partoens, F.M. Peeters, Wigner crystallization in transition metal dichalcogenides: A new approach to correlation energy, Physical Review B, 95 (2017) 115438.

[23] E. Alves, G.L. Bendazzoli, S. Evangelisti, J.A. Berger, Accurate ground-state energies of Wigner crystals from a simple real-space approach, Physical Review B, 103 (2021) 245125.

[24] T. Fujikawa, K. Hatada, L. Hedin, Self-consistent optical potential for atoms in solids at intermediate and high energies, Physical Review B, 62 (2000) 5387-5398.

[25] A.N. Kravtsova, A.V. Soldatov, A.M. Walker, A.J. Berry, The Ti environment in natural hibonite: XANES spectroscopy and computer modelling, Journal of Physics: Conference Series, 712 (2016) 012089.

# Supporting materials of BBC model

## BBC Model

**1. Binding energy shift (BB model)**      **2. Deformation bond energy (BC model)**

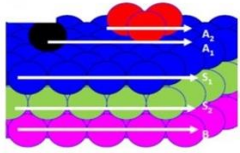 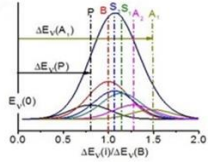 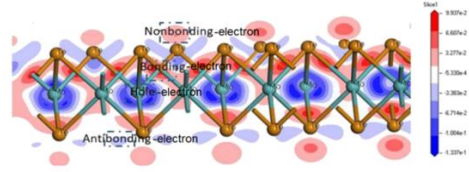

**Fig. S1** Schematic diagram of the BBC model.

The BBC model is quantified chemical bonds by binding energy shift and deformation charge density. For the binding energy (BB) model, we use the central field approximation and Tight-binding (TB) model, which can get the relationship between the energy shift and Hamiltonian. For the bond-charge (BC) model, we use the second-order term of energy expansion and use deformation charge density to calculate bonding states.

For binding energy (BB) model:

$$\begin{cases} H = \int \Psi^+(\vec{r}) h(\vec{r}) \Psi(\vec{r}) \mathrm{d}r = \sum_{l,l'} C_l^+ C_{l'} \int a^*(\vec{r}-\vec{l}) h(\vec{r}) a(\vec{r}-\vec{l'}) \mathrm{d}r \\ H = \xi(0) \sum_l \hat{n}_l - J \sum_l \sum_\rho C_l^+ C_{l+\rho} \\ J = \sum_{l,l'} C_l^+ C_{l'} \int a^*(\vec{r}-\vec{l}) \left[ V(\vec{r}) - v_a(\vec{r}-\vec{l}) \right] a(\vec{r}-\vec{l'}) \mathrm{d}r \\ \sum_l \hat{n}_l = C_l^+ C_l \end{cases}$$

(1)

$V(r)$ is the potential and $v_a(\vec{r}-\vec{l})$ is the atomic potential. $\xi(0)$ is the local orbital electron energy. $\hat{n}_l$ represents the electron number operator in the Wanier representation.

$$\begin{cases} \xi(0) = \left(E_n^a - A_n\right) \\ A_n = -\int a_n^*(\vec{r}-\vec{l})\left[V(\vec{r})-v_a(\vec{r}-\vec{l})\right]a_n(\vec{r}-\vec{l})\mathrm{d}r \\ E_n^a = \int a_n^*(\vec{r}-\vec{l})\left[-\frac{\hbar^2}{2m}\nabla^2 + v_a(\vec{r}-\vec{l})\right]a_n(\vec{r}-\vec{l})\mathrm{d}r \\ \Psi(\vec{r}) = \sum_l C_l a(\vec{r}-\vec{l}) \end{cases}$$

(2)

$E_n^a$ is the atomic energy level, and J is called the overlapping integral. $A_n$ represents the energy level shift, $E_n^a$ caused by the potential of (N-1) atoms outside the lattice point $l$ in the lattice

Consider the effect of external fields on energy level shifts:

$$\begin{cases} H' = \xi(0)\sum_l \hat{n}_l - J\sum_l\sum_\rho C_l^+ C_{l+\rho} = \left(E_n^a - \gamma A_n\right)\sum_l \hat{n}_l - \gamma J\sum_l\sum_\rho C_l^+ C_{l+\rho} \\ \Delta E_v(B) = A_n\sum_l \hat{n}_l, \quad \Delta E_v(x) = \gamma A_n \sum_l \hat{n}_l \\ V_{cry}(\vec{r}_i - \vec{l}_j) = V(\vec{r}) - v_a(\vec{r}-\vec{l}) \\ \gamma V_{cry}(\vec{r}_i - \vec{l}_j) = \gamma \sum_{i,j,l_j \neq 0} \frac{1}{4\pi\varepsilon_0}\frac{Z'e^2}{|\vec{r}_i - \vec{l}_j|} = \gamma \sum_{i,j,l_j \neq 0} \frac{1}{4\pi\varepsilon_0}\frac{(Z-\sigma_v)e^2}{|\vec{r}_i - \vec{l}_j|} \\ \Delta E_v(x) = \gamma \int a_n^*(\vec{r}-\vec{l}) \sum_{i,j,l_j \neq 0} \frac{1}{4\pi\varepsilon_0}\frac{(Z-\sigma_v)e^2}{|\vec{r}_i - \vec{l}_j|} a_n(\vec{r}-\vec{l})\mathrm{d}r \sum_l \hat{n}_l \\ \Delta E_v(x) = \gamma \left\langle \Psi_v(\vec{r}) \right| \sum_{i,j,\vec{R}_j \neq 0} \frac{1}{4\pi\varepsilon_0}\frac{(Z-\sigma_v)e^2}{|\vec{r}_i - \vec{l}_j|} \left| \Psi_v(\vec{r}) \right\rangle \end{cases}$$

(3)

The effective positive charge of the ion is $Z' = Z - \sigma_v$, considering the charge shielding effect $\sigma_v$, where Z is the nuclear charge. $\Delta E_v(B)$ is represents the energy shift of an atom in an ideal bulk. $\delta\gamma = \gamma - 1$ is relative bond energy ratio and B indicates bulk atoms.

For bond-charge (BC) model, we consider the positive charge background (*b*) and the electron (*e*) as a system, and write their Hamiltonian sums and their interactions,

respectively. In addition to electron kinetic energy, only electrostatic Coulomb interactions are considered:

$$\begin{cases} H = H_b + H_e + H_{eb} \\ H_e = \sum_{i=1}^{N} \frac{P_i^2}{2m} + \frac{1}{2}e_1^2 \sum_{i=1}^{N}\sum_{\substack{j=1 \\ i \neq j}}^{N} \frac{1}{|\vec{r}_i - \vec{r}_j|} e^{-\mu|\vec{r}_i - \vec{r}_j|} \\ H_b = \frac{1}{2}e_1^2 \int d^3x \int d^3x' \frac{n(\vec{x})n(\vec{x}')}{|\vec{x} - \vec{x}'|} e^{-\mu|\vec{x} - \vec{x}'|} = \frac{1}{2}e_1^2 \left(\frac{N}{V}\right)^2 \int d^3x 4\pi \int dz \frac{e^{-\mu z}}{z} = 4\pi e_1^2 \frac{N^2}{2V\mu^2} \\ H_{eb} = -e_1^2 \sum_{i=1}^{N} \int d^3x \frac{n(\vec{x})}{|\vec{x} - \vec{r}_i|} e^{-\mu|\vec{x} - \vec{r}_i|} = -e_1^2 \sum_{i=1}^{N} \frac{N}{V} 4\pi \int dz \frac{e^{-\mu z}}{z} = -4\pi e_1^2 \frac{N^2}{V\mu^2} \end{cases}$$

(4)

The shielding factor $\mu$ is added to the equation. $\vec{r}_i$ represents the *ith* electronic position. $\vec{x}$ represents the background position. $e$ is the basic charge, $e_1 = e/\sqrt{4\pi\varepsilon_0}$. $e_1 n(\vec{x})$ is the charge density at background $\vec{x}$, and $n(\vec{x}) = N/V$ is a constant.

$$\begin{cases} H_e = \sum_{i=1}^{N} \frac{P_i^2}{2m} + \frac{1}{2}e_1^2 \sum_{i=1}^{N}\sum_{\substack{j=1 \\ i \neq j}}^{N} \frac{1}{|\vec{r}_i - \vec{r}_j|} e^{-\mu|\vec{r}_i - \vec{r}_j|} \\ = \sum_{k\sigma} \frac{\hbar^2 k^2}{2m} a_{k\sigma}^\dagger a_{k\sigma} + \frac{e_1^2}{2V} \sum_{\vec{q}}^{*} \sum_{\vec{k}\sigma} \sum_{\vec{k}'\lambda} \frac{4\pi}{q^2 + \mu^2} a_{\vec{k}+\vec{q},\sigma}^\dagger a_{\vec{k}'-\vec{q},\lambda}^\dagger a_{\vec{k}'\lambda} a_{\vec{k}\sigma} + \frac{e_1^2}{2V}\frac{4\pi}{\mu^2}(N^2 - N) \\ \sum_{i=1}^{N} \frac{P_i^2}{2m} = \sum_{l'\sigma'}\sum_{l\sigma} a_{\vec{k}_{l'}\sigma'}^\dagger \left\langle \vec{k}_{l'}\sigma' \left| \frac{P^2}{2m} \right| \vec{k}_l\sigma \right\rangle a_{\vec{k}_l\sigma} = \sum_{l\sigma} \frac{\hbar^2 k_l^2}{2m} a_{\vec{k}_l\sigma}^\dagger a_{\vec{k}_l\sigma} \\ \frac{1}{2}e_1^2 \sum_{i=1}^{N}\sum_{\substack{j=1 \\ i \neq j}}^{N} \frac{1}{|\vec{r}_i - \vec{r}_j|} e^{-\mu|\vec{r}_i - \vec{r}_j|} = \frac{1}{2}e_1^2 \sum_{l'\sigma'}\sum_{m'\lambda'}\sum_{l\sigma}\sum_{m\lambda} a_{\vec{k}_{l'}\sigma'}^? a_{\vec{k}_{m'}\lambda'}^\dagger \left\langle \vec{k}_{l'}\sigma', \vec{k}_{m'}\lambda' \left| \frac{e^{-\mu|\vec{r}_i - \vec{r}_j|}}{|\vec{r}_i - \vec{r}_j|} \right| \vec{k}_l\sigma, \vec{k}_m\lambda \right\rangle a_{\vec{k}_m\lambda} a_{\vec{k}_l\sigma} \\ = \frac{1}{2}e_1^2 \sum_{l'}\sum_{m'}\sum_{l\sigma}\sum_{m\lambda} \delta_{\sigma\sigma'}\delta_{\lambda\lambda'} a_{\vec{k}_{l'}}^? a_{\vec{k}_{m'}}^\dagger \left\langle \vec{k}_{l'}\vec{k}_{m'} \left| \frac{e^{-\mu|\vec{r}_i - \vec{r}_j|}}{|\vec{r}_i - \vec{r}_j|} \right| \vec{k}_l\vec{k}_m \right\rangle a_{\vec{k}_m} a_{\vec{k}_l} \\ = \frac{e_1^2}{2V} \sum_{\vec{k}}\sum_{\vec{k}'}\sum_{\vec{q}}\sum_{\sigma\lambda} \frac{4\pi}{q^2 + \mu^2} a_{\vec{k}+\vec{q},\sigma}^\dagger a_{\vec{k}'-\vec{q},\lambda}^\dagger a_{\vec{k}'\lambda} a_{\vec{k}\sigma} \end{cases}$$

(5)

In the formula, the $q$ on the sum sign of "*" indicates that the part where $q = 0$ is ignored during the sum. When $V \to \infty, N \to \infty$, while keeping $N/V$ constant, the last term to the right of the equal sign causes the average energy $H_e/N$ of each particle

to become:

$$\frac{1}{2}4\pi e_1^2 \left(\frac{N}{V}\right)\frac{1}{\mu^2} - \frac{1}{2}4\pi e_1^2 \left(\frac{N}{V}\right)\frac{1}{N}\frac{1}{\mu^2}$$

The former term is constant, and the latter term tends to zero. If $\mu \to 0$, the former term becomes a divergent term. However, this term just cancels out the divergent $H_b$ and $H_{eb}$ term. Thus, the Hamiltonian of the system becomes

$$H = \sum_{\vec{k}\sigma} \frac{\hbar^2 \vec{k}^2}{2m} a^\dagger_{\vec{k}\sigma} a_{\vec{k}\sigma} + \frac{e_1^2}{2V} \sum_{\vec{q}}{}^* \sum_{\vec{k}\sigma} \sum_{\vec{k}'\lambda} \frac{4\pi}{q^2} a^\dagger_{\vec{k}+\vec{q},\sigma} a^\dagger_{\vec{k}'-\vec{q},\lambda} a_{\vec{k}'\lambda} a_{\vec{k}\sigma}$$

(6)

Electron interactions expressed using electron density:

$$\begin{cases} \hat{V}_{ee} = \dfrac{e_1^2}{2V} \sum_q {}^* \sum_{\vec{k}\sigma} \sum_{\vec{k}'\lambda} \dfrac{4\pi}{q^2} a^\dagger_{\vec{k}+\vec{q},\sigma} a^\dagger_{\vec{k}'-\vec{q},\lambda} a_{\vec{k}'\lambda} a_{\vec{k}\sigma} \\[6pt] = \dfrac{1}{2} e_1^2 \sum_{l'\sigma'} \sum_{m'\lambda'} \sum_{l\sigma} \sum_{m\lambda} a^?_{\vec{k}_{l'}\sigma'} a^\dagger_{\vec{k}_{m'}\lambda'} \left\langle \vec{k}_{l'}\sigma', \vec{k}_{m'}\lambda' \left| \dfrac{1}{\|\vec{r}_i - \vec{r}_j\|} \right| \vec{k}_l\sigma, \vec{k}_m\lambda \right\rangle a_{\vec{k}_m\lambda} a_{\vec{k}_l\sigma} \\[6pt] = \dfrac{1}{2} \sum_{\vec{k}_1\vec{k}_2, \vec{k}_1'\vec{k}_2'} \sum_{\sigma_1\sigma_2} \left\langle \vec{k}_1, \vec{k}_2 |v| \vec{k}_1', \vec{k}_2' \right\rangle C^+_{\vec{k}_1\sigma_1} C^+_{\vec{k}_2\sigma_2} C_{\vec{k}_2'\sigma_2} C_{\vec{k}_1'\sigma_1} = \dfrac{1}{2} U \sum_i \sum_{\sigma\sigma'} C^+_{i\sigma} C^+_{i\sigma'} C_{i\sigma'} C_{i\sigma} = = \dfrac{1}{2} U \sum_i \sum_{\sigma\sigma'} n_{i\sigma} n_{i\bar\sigma} \\[6pt] = \dfrac{1}{4\pi\varepsilon_0} \times \dfrac{1}{2|\vec{r}-\vec{r}'|} \int d^3r \int d^3r' \rho(\vec{r}) \rho(\vec{r}') \\[6pt] \rho(\vec{r}) = \alpha_\zeta^+(\vec{r}) \alpha_\zeta^-(\vec{r}), \alpha_\zeta^+(\vec{r}) = \sum_{\vec{k}} \psi^*_{\vec{k}}(\vec{r}) a^+_{\vec{k}\zeta} \equiv \sum_i \psi^*_{\vec{k}i}(\vec{r}) a^+_{i\zeta} \\[6pt] \left\langle k_1, k_2 |v| k_1', k_2' \right\rangle = \int \dfrac{e^2 \psi^*_{k_1}(\vec{r}) \psi^*_{k_2}(\vec{r}') \psi_{k_1'}(\vec{r}) \psi_{k_2'}(\vec{r}')}{4\pi\varepsilon_0 |\vec{r}-\vec{r}'|} drdr' \end{cases}$$

(7)

Electron interaction terms for density fluctuations:

$$\begin{cases} \delta V_{ee} = \dfrac{1}{4\pi\varepsilon_0} \times \dfrac{1}{2|\vec{r}-\vec{r}'|} \int d^3r \int d^3r' \delta\rho(\vec{r}) \delta\rho(\vec{r}') \\[6pt] = \pm \dfrac{1}{4\pi\varepsilon_0} \times \dfrac{1}{2|\vec{r}-\vec{r}'|} \int d^3r \int d^3r' \rho(\vec{r}) \rho(\vec{r}') e^{-\mu(\vec{r}-\vec{r}')} \\[6pt] = \pm \dfrac{e_1^2}{2V} \sum_{\vec{k}} \sum_{\vec{k}'} \sum_{\vec{q}} \sum_{\sigma\lambda} \dfrac{4\pi}{q^2+\mu^2} a^\dagger_{\vec{k}+\vec{q},\sigma} a^\dagger_{\vec{k}'-\vec{q},\lambda} a_{\vec{k}'\lambda} a_{\vec{k}\sigma} \\[6pt] \Delta V_{bc}(\vec{r}-\vec{r}') = \delta V_{ee} = \dfrac{1}{4\pi\varepsilon_0} \dfrac{1}{2|\vec{r}-\vec{r}'|} \int d^3r \int d^3r' \delta\rho(\vec{r}) \delta\rho(\vec{r}') \end{cases}$$

(8)